\documentclass{article}

\usepackage[sort&compress,numbers]{natbib}
\usepackage{graphicx}%
\usepackage{multirow}%
\usepackage{amsmath,amssymb,amsfonts,geometry}%
\usepackage{amsthm}%
\usepackage{mathrsfs}%
\usepackage[title]{appendix}%
\usepackage{xcolor}%
\usepackage{textcomp}%
\usepackage{manyfoot}%
\usepackage{booktabs}%
\usepackage{algorithm}%
\usepackage{algorithmic}%
\usepackage{listings}%
\usepackage{ulem}
\usepackage{subfigure}
\usepackage{caption}
\usepackage{indentfirst}
\usepackage{url}
\usepackage[thicklines]{cancel}
\usepackage{doi}
\usepackage{hyperref}

\usepackage{xcolor}

\let\svthefootnote\thefootnote
\newcommand\freefootnote[1]{%
  \let\thefootnote\relax%
  \footnotetext{#1}%
  \let\thefootnote\svthefootnote%
}

\graphicspath{ {./Images/} }

\title{EdSr: A Novel End-to-End Approach for State-Space Sampling in Molecular Dynamics Simulation}

\author{Hai-Ming Cao$^1$, Bin Li$^1$*}
\begin{document}
\date{}
\maketitle

\freefootnote{$^1$School of Chemical Engineering and Technology, Sun Yat-Sen University, Zhuhai 519082, China.
E-mail: libin76@mail.sysu.edu.cn}

\section*{\centering Abstract}\label{abstract}

The molecular dynamics (MD) simulation technique has been widely used in complex systems, but the accessible time scale is limited due to the requirement of small integration timesteps.
  Here, we propose a novel method, named Exploratory dynamics Sampling with recursion (EdSr), 
 inspired by ordinary differential equation and Taylor expansion formula, which enables flexible  adjustment of timestep in MD simulations.
  By setting up four groups of experiments including simple functions, ideal physical models, all-atom simulations and coarse-grained simulations, 
 we demonstrate that EdSr can dynamically and flexibly adjust the simulation timestep according to the requirements during simulation period, and operate with larger timesteps than the widely used velocity-Verlet integrator.
  Although the method can not perform perfectly at flexible timestep across all simulation systems,
 we believe that it provides a promising direction for future studies.

\section{Introduction}\label{introduction-edsr-exploratory-dynamics-sampling-with-recursion}

The molecular dynamics (MD) simulation method, a branch of the multiscale modeling architecture, has been widely used in many research areas,
such as soft matter, chemical engineering, material science, and
biological systems.
Rooted in Newton's second law of motion and statistical mechanics, MD simulation method was firstly developed in the 1950s \cite{alder_phase_1957}.
Nowadays, the efficiency and accuracy of MD simulations have been improved significantly due to enhancements of computer hardware and algorithms.
A series of MD simulation techniques have been developed to address different time and lengths scales \cite{muller-plathe_coarse-graining_2002,peter_multiscale_2009}. 
For instance,  all-atom (AA) MD simulation is able to
simulate evolution of complicated systems with a relatively high precision \cite{allen_computer_1987,Zeolite,AAMD_membrane_fusion}; 
while coarse-grained (CG) simulation method, which treats several atoms or molecules into a single CG bead, captures the global behavior of macromolecules at lower
computational cost with an acceptable bias \cite{kmiecik_coarse-grained_2016,marrink_martini_2007,brini_systematic_2013,CGmartini,CGmartini2,CGmartini3}.

Although MD simulations exhibit many advantages, like the investigation of underlying mechanism in experimental phenomenons, prediction for future material designs, and so on,
the computation of complex interactions, the limitation of
timestep \(\Delta t\), and other factors result in expensive computational cost in MD simulations.
Some researchers have
attempted to reduce the computational cost through  force field fitting,
such as Machine learning Force Field \cite{wang_enhancing_2024,schutt2021equivariant,wieser2024machine}. 
However, the integration timestep still limits the accessible time scale of MD simulation.
Some of the classical time
integration algorithms, such as velocity-Verlet integrator \cite{VerletVelocity}, predictor-corrector \cite{PredictorCorrector}, leap frog
integrator, multi-stepping time integrator \cite{rRESPA}, have been practically implemented to solve evolution of complicated molecular systems in some conventional simulation software, like LAMMPS \cite{LAMMPS_citation}, GROMACS \cite{cite_gromacs_1,cite_groamcs_2}, etc.
Meanwhile,  methods like Physics-Informed Neural Networks (PINN) \cite{cranmer2020lagrangian,lutter2018deep,greydanus2019hamiltonian,sosanya2022dissipative}, combining machine learning and physical mechanics, 
have been proposed for solving ordinary differential equation (ODE) with respect to evolution of dynamical systems.

Recently, with the development of machine learning, as well as the success of
score-based model based on Langevin dynamics \cite{song2021scorebasedgenerativemodelingstochastic} and Denoising Diffusion Probabilistic
Model (DDPM) \cite{pmlr-v37-sohl-dickstein15,ho_denoising_2020} in
the field of Computer Vision (CV), researchers have adopted these methods
for enlarging timestep in MD simulation, and have proposed a series of methods on
the basis of diffusion model. For example, Wu et al. proposed an
equivariant geometric transformer based on diffusion process to sample atomic
positions \cite{wu2023diffmd,lee_score-based_2023}; Schreiner et al. introduced an Implicit Transfer Operator
(ITO) for building multiple time-scale surrogate models \cite{schreiner2024implicit}; 
Hsu and co-workers adopted score dynamics to accelerate the evolution of MD simulations \cite{hsu_score_2024}.

While the aforementioned methods have been able to sample state of ensembles at larger timesteps
by predicting probability density distributions, there are still some challenges remained. 
Firstly, timestep adjustment is often inflexible, 
as these methods only sample positions at fixed intervals  
\(\Delta t\) identical to those used in training. On the other hand, some of these methods need extensive datasets to train machine learning models.
To address the issues,  we propose a novel algorithm named
\textbf{Exploratory dynamics Sampling with recursion (EdSr)},  based on Taylor expansion and ODE. 
By using a recursive process, the aim of our method is to
adjust the simulation timestep  dynamically and flexibly according to requirements during
simulation period without relying on data-driven training.

In order to evaluate the performance and accuracy of the EdSr method, we conduct comprehensive
experiments on simple functions, ideal physical
models \cite{greydanus2019hamiltonian}, atomistic simulations \cite{Zeolite}, and
coarse-grained simulations \cite{morton_specificity_1995,lee_coarse-grained_2009,ubiquitin}. By comparing
with  velocity-Verlet (VV) integrator (one of the most used methods in MD simulations),
the EdSr method not only achieves higher precision at the same timestep, but
also is able to maintain acceptable accuracy at timesteps at least twice as large (depending on the complexity of
explicit ensemble).
In addition, we also test negative integration timesteps for ideal physical models, to validate the time reversible symmetry of EdSr method,
which demonstrates  a high
precision of EdSr for the ability of time reversible symmetry.

\section{Methods}\label{method}

\subsection{Preliminary}\label{preliminary}

\textbf{Path Integration} 
\\A time-dependent variable $X(t)$ can be described by
the differential equation \(\displaystyle X'(t) = \frac{dX}{dt} = f(t)\) for a
general time \(t\). Given an interval \(\Delta t\), \(X_{t + \Delta t}\)
is given by:

\begin{equation} \label{eq:path_integration}
  X_{t + \Delta t} = X_t + \int^{t + \Delta t}_{t} f(t') dt'  
\end{equation}

\textbf{Taylor Expansion} 
\\For a real or complex
function \(f(x)\), that is infinitely differentiable at a real or
complex number \(x_0\), the Taylor expansion at \(x_0 + \Delta x\) is:

\begin{equation} \label{eq:taylor_expansion}
  f(x_0 + \Delta x) = f(x_0)+{\frac {f'(x_0)}{1!}}\Delta x+{\frac {f''(x_0)}{2!}}\Delta x^{2}+\cdots+ {\frac {f^{(n)}(x_0)}{n!}}\Delta x^{n} 
\end{equation}

where \(n\to\infty\) and \(\Delta x\) is within domain of convergence,
namely that \(|\Delta x|\) lies within the radius of convergence \(r\).

\subsection{Exploratory dynamics Sampling with recursion}

EdSr is based on the following hypotheses:

\begin{enumerate}
\def\labelenumi{\arabic{enumi}.}
\item \label{hypo:1}
  The time evolution function \(X(t)\) is reasonable and accessible, which
  corresponds to an actual state of a physical system.
  \(X(t)\) is a continuous function in solving interval.
\item \label{hypo:2}
   \(X(t)\) is infinitely differentiable at time \(t\) for an arbitrary physics system, namely that
  \(\displaystyle \lim_{n \rightarrow +\infty} X^{(n)}(t) \neq \infty\).
\item \label{hypo:3}
  The differential operator \(D\) satisfies:
    \begin{align*}
    D_t(x(t) + y(t)) ={}  D_t x(t) + D_t y(t), \qquad  D_t (D_t x(t)) = D^2_t x(t) \qquad
    D_t (ax(t)) ={}  a D_t x(t) = D_t x(t) \cdot a
    \end{align*}
    where $a$ is a constant.
\item \label{hypo:4}
  \(X^{(n-1)}(t)\) and \(X^{(n)}(t)\) satisfy the following inequality:
    \begin{equation}
    \lim_{n \rightarrow +\infty} \frac{X^{(n)}(t)}{nX^{(n-1)}(t)} \neq 0 \notag
    \end{equation}
\end{enumerate}

Given an initial state \(X_t\), 
the corresponding force and the next state \(X_{t + \Delta t}\) can be obtained through the Newton's second law of motion (\(F = ma\)),
and the equation of motion (\(\displaystyle X_{t + \Delta t} = X_t + v_t \Delta t + \frac{1}{2} \frac{F}{m} \Delta t^2\)), respectively.
  However, it is obvious that \(\Delta t\) should be small enough to maintain an acceptable bias.
For a specific physical system,
 if the function \(f(t)\) in Equation (\ref{eq:path_integration}) was known, 
 the state \(X_{t + \Delta t}\) after a time interval \(\Delta t\) can be solved  using path integration, 
 thus alleviating the constraint of small \(\Delta t\). 
  Therefore, we are going to consider to solve the path integration.
However, it is difficult to get the \(X'(t) = f(t)\) over time \(t\) in general.
  So we use partial integration to transform path integration formula, which is described as:
\begin{equation} \label{eq:path_integration_expansion}
X(b) = X(a) + X'(a)\Delta t + \int^{\Delta t}_{0} yX''(b - y)dy
\end{equation}
where \(b\) and \(a\) are equivalent to \(t + \Delta t\) and \(t\), respectively.
  According to First Mean Value Theorems for Definite Integrals,
  Equation (\ref{eq:path_integration_expansion}) is transformed to the following form:
\begin{equation} \label{eq:integration_mean_value}
X(b) = X(a) + X'(a)\Delta t + \frac{1}{2} X''(c) \Delta t^2 \qquad c \in (a, b)
\end{equation}
Equation (\ref{eq:integration_mean_value}) indicates that 
 \(X_{t + \Delta t}\) can be obtained from the \(X_t\) (generalized coordinates), 
 the corresponding first order derivative \(X'(t)\) (generalized velocities), and the second order derivative of an intermediate point \(c\) between \(t\) and \(t + \Delta t\), at a specific interval \(\Delta t\).
  Consequently, the aim to solve the \(f(t)\) transforms to the aim to solve the point \(c\).

According to the hypothesis (\ref{hypo:2}),
  Equation (\ref{eq:path_integration_expansion}) can be expanded using \(N\)-times  partial integration and we replace the derivative, $X(a)$ by the differential operator \(D\), \(X_N\) respectively.
Then we obtain:

\begin{equation} \label{eq:terms}
  \begin{aligned}
    X(b) ={} & X_N + DX_N\Delta t + \underbrace{\frac{1}{2!} D^2X_N\Delta t^2 + ... + \frac{1}{N!} D^NX_N\Delta t^N}_{\rm{\mathbf{the\ first\ term}}} \\
    & + \underbrace{\frac{1}{(N+1)!}\int^{\Delta t}_{0} X^{(N+1)}(b - y)dy^{N + 1}}_{\rm{\mathbf{the\ second\ term}}}
  \end{aligned}
\end{equation}
For the last three terms in ``the first term'' in Equation (\ref{eq:terms}),  we can merge them into one term according to hypothesis (\ref{hypo:3}):
\begin{align}
  & \frac{1}{(N-2)!} D^{N-2}X_N\Delta t^{N-2} + \frac{1}{(N-1)!} D^{N-1}X_N\Delta t^{N-1} + \frac{1}{N!} D^NX_N\Delta t^N  \\
  ={} & \frac{1}{(N-2)!} D^{N-2}\bigg(X_N + \frac{1}{N-1}DX_N\Delta t + \frac{1}{(N-1)N}D^2X_N\Delta t^{2}\bigg)\Delta t^{N-2} \\
  ={} & \frac{1}{(N-2)!} D^{N-2}\Bigg(X_N + \frac{1}{N-1} \Big(DX_N\Delta t + \frac{1}{N}D^2X_N\Delta t^{2}\Big)\Bigg)\Delta t^{N-2} \\
  ={} & \frac{1}{(N-2)!} D^{N-2} \underline{\mathbf{X_{N-2}}} \Delta t^{N-2},\qquad \mathrm{where}\ \mathbf{X_{N-2}} = X_N + \frac{1}{N-1} \Big(DX_N\Delta t + \frac{1}{N}D^2X_N\Delta t^{2}\Big)
\end{align}
Analogously, we can perform the  operation again:
\begin{align}
  & \frac{1}{(N-4)!} D^{N-4}X_N\Delta t^{N-4} + \frac{1}{(N-3)!} D^{N-3}X_N\Delta t^{N-3} + \frac{1}{(N-2)!} D^{N-2}\mathbf{X_{N-2}}\Delta t^{N-2}  \\
  ={} & \frac{1}{(N-4)!} D^{N-4}\Bigg(X_N + \frac{1}{N-3} \Big(DX_N\Delta t + \frac{1}{N-2}D^2\mathbf{X_{N-2}}\Delta t^{2}\Big)\Bigg)\Delta t^{N-4} \\
  ={} & \frac{1}{(N-4)!} D^{N-4} \underline{\mathbf{X_{N-4}}} \Delta t^{N-4}, \qquad \mathrm{where}\ \mathbf{X_{N-4}} = X_N + \frac{1}{N-3} \Big(DX_N\Delta t + \frac{1}{N-2}D^2\mathbf{X_{N-2}}\Delta t^{2}\Big)
\end{align}
After the process is executed by \(\displaystyle \frac{N}{2}\) times, the whole ``the first term'' in Equation \ref{eq:terms} will be ``merged'' into one term \(\mathbf{X_0}\), 
  derived from \(\displaystyle X_N + DX_N\Delta t + \frac{1}{2!} D^2\mathbf{X_2}\Delta t^2\). 
  For ``the second term'' in Equation (\ref{eq:terms}), because \(X(b)\) exists according to hypothesis (\ref{hypo:1}), 
 the series is convergent. 
  Thereby, ``the second term'' tends to be 0 as \(n \rightarrow \infty\),
  so we assume it to be 0. 
  According to Newton's second law of motion,
 \(D^2\mathbf{X_N}\) can be replaced by  \(\displaystyle -\frac{\nabla_X U(\mathbf{X_N})}{M} \),  and we approximate \(X''(c)\) in Equation (\ref{eq:integration_mean_value}) via recursion.
Eventually, the part of displacement of EdSr can be rewritten as the following form:

\begin{equation}
X_{n-1} = X_N + \frac{1}{2n-1} \Big(X'_N\Delta t - \frac{1}{2n}\frac{\nabla_X U(X_n) }{M} \Delta t^{2}\Big), \quad n\ =\ N\ \mathrm{to}\ 1 \label{eq:EdSr_displacement}
\end{equation}

where \(X_0\), \(X_N\), \(X_N'\) denote \(X(b)\), \(X(a)\), \(X'(a)\), respectively.
We exhibit some examples for the derivations in Supporting Information Section 1.1.
According to the definition of derivative,
 the part of velocity of EdSr can be expressed as:

\begin{equation}
\left\{
\begin{aligned}
X_{n-1} ={} & X_N + \frac{1}{2n-2} \Big(X'_N\Delta t - \frac{1}{2n-1}\frac{\nabla_X U(X_n)}{M}\Delta t^{2}\Big), && \quad n\ =\ N\ \rm{to}\ 2 \\ \label{eq:EdSr_velocity}
X'_0 ={} & X'_N - \frac{\nabla_X U(X_1) }{M}\Delta t,   && \quad  n\  =\  1
\end{aligned}
\right.
\end{equation}

where \(X'_0\) denotes \(X'(b)\). 
The full details for the derivation of velocity of EdSr are provided in Section 1.2 of Supporting Information.
  The framework of the EdSr is summarized in Algorithm 1. 
In addition, we establish a connection between EdSr and Markov process, the discussions are shown in Supporting Information Section 2.

\begin{algorithm}[!htb]
\caption{Exploratorily dynamics Sampling with recursion}\label{ErS}
\hspace*{0.0cm}{\bf Input:}
Initial state $x_t = X(t)$, $v_t = X'(t)$, timestep $\Delta t$, mass $M$, force field solver $F(\cdot)$. \\
Max Iteration $N$;
\begin{algorithmic}[1]
\STATE{\bf Initialize:} $x_N \leftarrow x_t$, $y_N \leftarrow x_t$
\STATE{} $//$ Displacement Iteration;
\STATE\ $//$ \(x_n\), \(dx\), \(x_t\): \textbf{length} unit (meter or nm or \AA, etc)
\STATE\ $//$ \(v_t\): \textbf{velocity} unit (meter/s or nm/ps or \AA/fs, etc)
\FOR{$n := N$ to $1$}
\STATE $\displaystyle dx = v_t\ \Delta t + \frac{1}{2n}\ M^{-1}F(x_n)\ \Delta t^2$
\STATE $\displaystyle x_{n-1} = x_t + \frac{1}{2n - 1}\ dx$
\ENDFOR 
\STATE $//$ Velocity Iteration; 
\STATE $//$ \(y_n\), \(x_t\): \textbf{length} unit (meter or nm or \AA, etc)
\STATE $//$ \(v_t\), \(v_{t+ \Delta t}\), \(dy\): \textbf{velocity} unit (meter/s or nm/ps or \AA/fs, etc)
\FOR{$n := N$ to $2$}
\STATE $dy = \displaystyle \frac{1}{2n-1}\ M^{-1}F(y_n)\ \Delta t$
\STATE $y_{n-1} = x_t + \displaystyle \frac{1}{2n - 2}(v_t + dy)\ \Delta t$
\ENDFOR 
\STATE $v_{t + \Delta t}  = v_t + M^{-1}F(y_1)\ \Delta t$ 
\STATE $x_{t + \Delta t} = x_0$
\RETURN $x_{t + \Delta t}$, $v_{t + \Delta t}$
\end{algorithmic}
\end{algorithm}

\section{Results}\label{result}

In order to test the performance of the EdSr method, 
 we conduct comprehensive experiments with increasing complexity: simple functions, ideal physical models, all-atom MD simulations and coarse-grained MD simulations.

\subsection{Simple Function}\label{simple-function}

We design two groups of sub-experiments for simple function tests, in order to investigate the influence of the interval $\Delta x$ of EdSr on the reconstruction of function $f(x)$,  and the performance of constructing function by EdSr.
  The first group involves constructing functions with a fixed initial point \(x_0\) as a hyperparameter and a variable interval \(\Delta x\).
  The other group uses both initial point \(x_0\) and interval \(\Delta x\) as two fixed hyperparameters.
The experimental approach for performing the simple functions is described in Supporting Information Section 3.1.

\begin{figure}[h!]
  \captionsetup{justification=centering}
  \begin{minipage}[t]{0.5\textwidth}
      \centering
      \includegraphics[width=\textwidth]{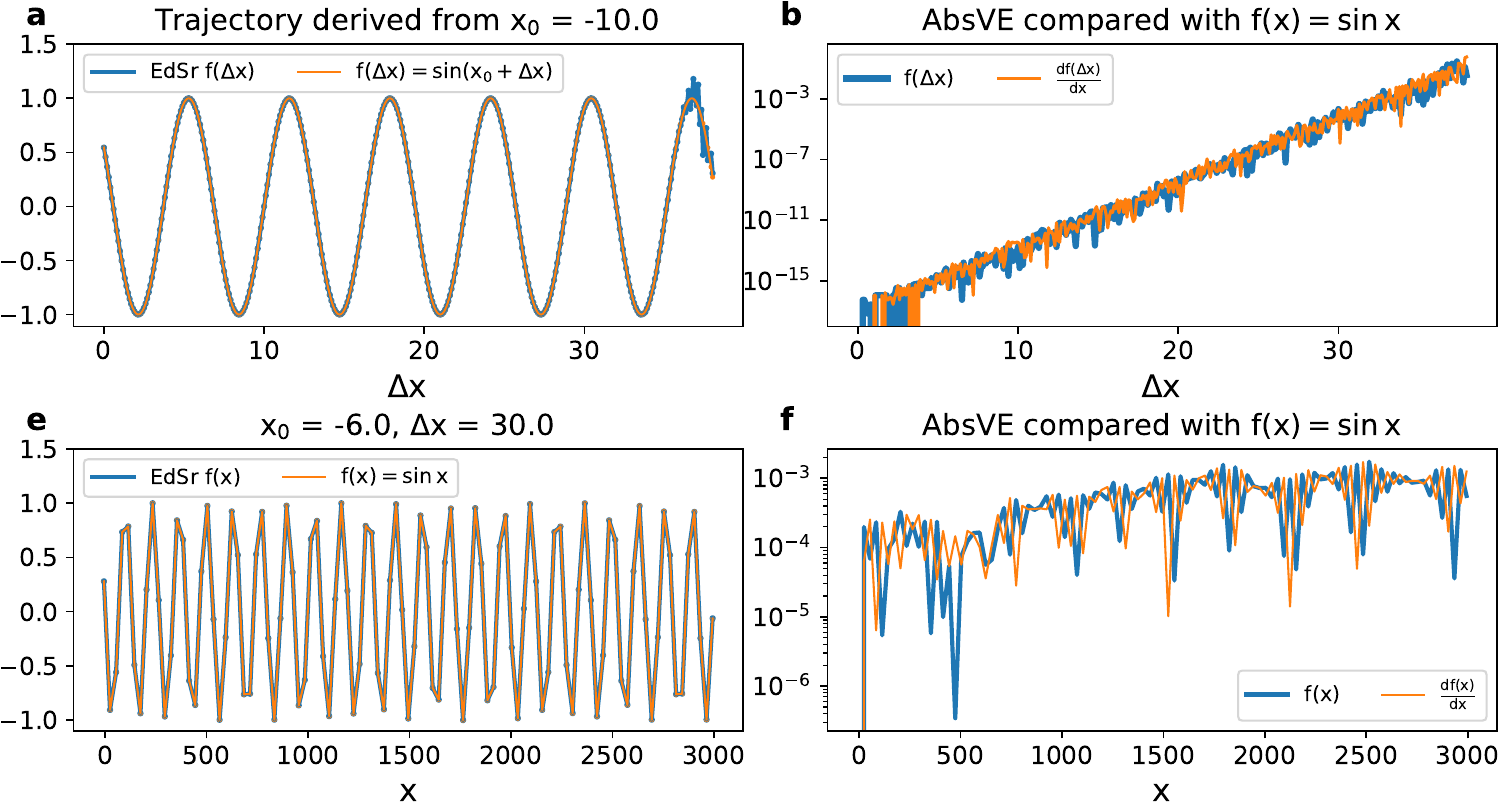}

  \end{minipage}
  \begin{minipage}[t]{0.5\textwidth}
      \centering
      \includegraphics[width=\textwidth]{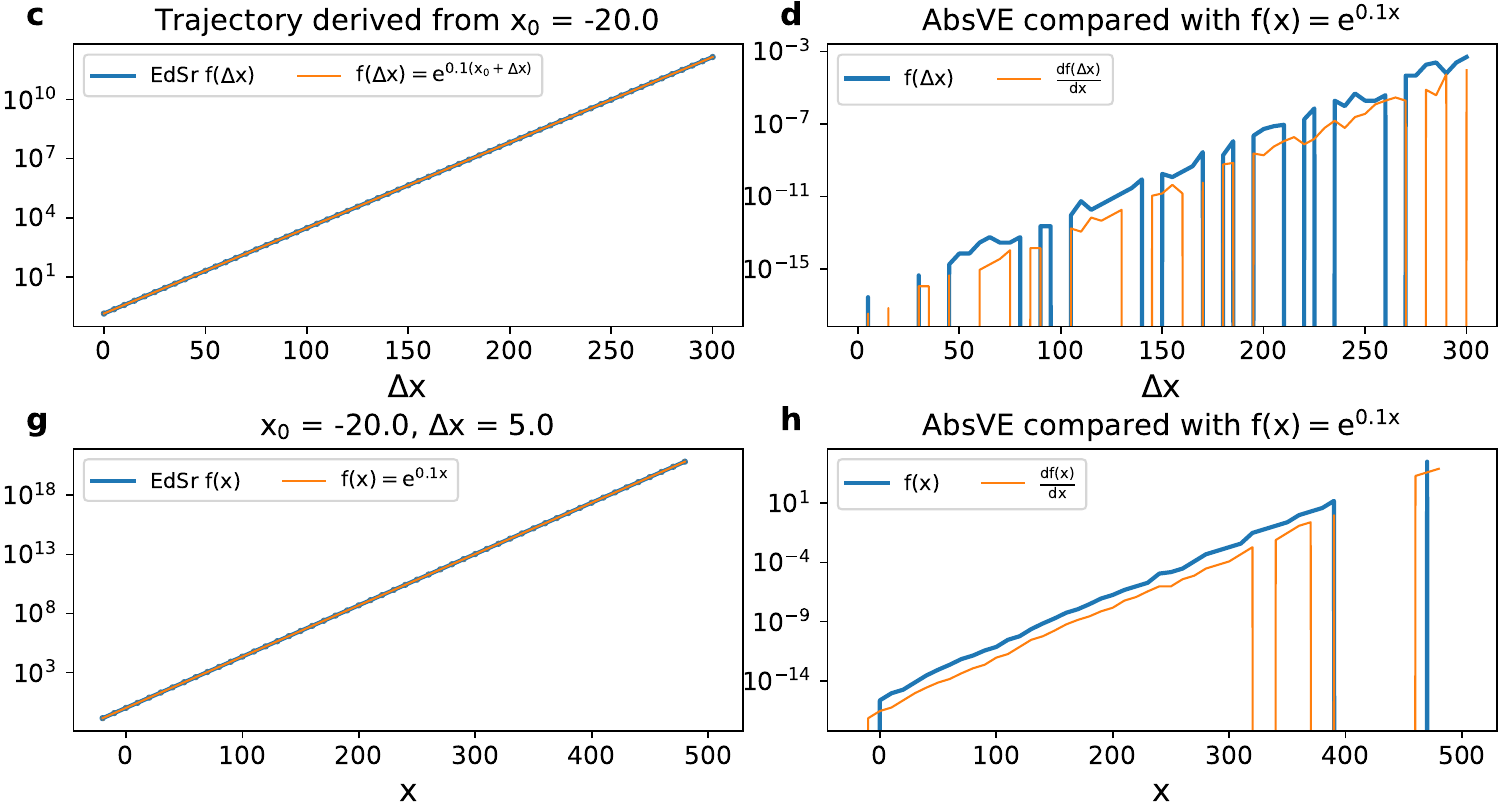}

  \end{minipage}
  \caption{
    Results of the two groups of sub-experiments of simple functions. 
    (a-d) show the results of the first group of sub-experiments, including the trajectories versus $\Delta x$, as well as the AbsVEs compared with source functions;
    (e-h) show the results of the second group of sub-experiments.
    (a),(b),(e),(f) show the results of $f(x) = \sin x$;
    (c),(d),(g),(h) show the results of $f(x) = e^{0.1x}$;
  }
  \label{picture:sinx_exp}
\end{figure}

Figure \ref{picture:sinx_exp}a and c show the comparisons between the two source functions ($f(x) = \sin x$ and $f(x) =  e^{0.1x}$) and the results constructed by EdSr at various $\Delta x$.
We observe excellent consistency between the source function of $f(x) = \sin x$ and the reconstruction of EdSr, significant deviation only emerges when the interval $\Delta x$ exceeds 35.0 (Figure \ref{picture:sinx_exp}a).
The results of $f(x) = e^{0.1x}$ obtained via EdSr approach show excellent consistency across all the tested intervals ($\Delta x$) (Figure \ref{picture:sinx_exp}c).
The absolute value of error (AbsVE) between the EdSr results and the source function, as well as their corresponding derivatives, are shown in Figure \ref{picture:sinx_exp}b and d.
Here, we define AbsVE as follows:
for the $n$-dimensional vectors $x$, $y$, the AbsVE is given by
\begin{equation}
  \mathrm{AbsVE}(x, y) = |x_i - y_i| \label{eq:definition_of_AbsVE}
\end{equation}
where \(x_i\), \(y_i\) denote the $i$-th elements of $x$, $y$, respectively.
Here, AbsVE reflects the absolute value of error between the EdSr reconstructions and the original functions.
  Although the AbsVE increases with the interval \(\Delta x\),
 the scales of AbsVE shown in Figure \ref{picture:sinx_exp}b and d indicate that EdSr maintains relatively high precision across different intervals.
In the second group of sub-experiments, 
 we fix both initial point \(x_0\) and interval \(\Delta x\), and set $f(x_0 + n  \Delta x)$ derives from $f(x_0 + (n-1)  \Delta x)$ to construct functions.
  The comparisons between the results of  EdSr and source function, as well as the AbsVEs and their corresponding derivatives are shown in Figure \ref{picture:sinx_exp}e-h. 
  EdSr also exhibits higher precision in a long series, as evidenced by the close match to the source functions and the corresponding derivatives.
  Note that as shown in Figure \ref{picture:sinx_exp}h, 
 although the scales of AbsVE results of both functions and derivatives exceed $10^1$,
 the deviations are negligible compared with the scale of the source function values (Figure \ref{picture:sinx_exp}g).
  In addition, we also evaluate the performance of EdSr with some other functions including \(f(x) = x^2 - 2x - 5\), \(f(x) = \mathrm{sigmoid}(x)\) and \(f(x) = x^3\). The comparisons between the results obtained via EdSr and the source functions  are shown from Figure S1 to Figure S4 in Supporting Information.

\subsection{Ideal Physical Models}\label{ideal-physical-models}
 
In this part, we refer to some ideal physical models, including ideal spring model, ideal pendulum model and two-body model, which were also studied in  Hamiltonian Neural Networks (HNN)\cite{greydanus2019hamiltonian}, and adopt them to verify the performance of EdSr.
  We also design the same two groups of sub-experiments, including a fixed initial time \(t_0\) and a variable interval \(\Delta t\),
  as well as both fixed \(t_0\) and interval \(\Delta t\),
 the setting is identical as in simple functions respectively.
   In order to check the time reversible symmetry of EdSr method on the ideal physical models, we set two series of $\Delta t$  with positive $\Delta t$ and negative $\Delta t$.
  Besides, we use Runge-Kutta method of order 5(4) (RK45)\cite{RungeKutta} and smaller step size to generate trajectory data as ground truth (GT), 
 the data generated by velocity-Verlet integrator (VV) are set as control group. 

\textbf{Ideal Spring}

First, we adopt the ideal spring model to test the performance of EdSr,
in which the Hamiltonian is:
\begin{equation}
\mathcal{H} = \frac{1}{2}kq^2 + \frac{1}{2}mv^2
\end{equation}
 where $k$ is the spring constant, $q$ is the generalized coordinate, and $m$ is the mass, $v$ is the velocity. Elasticity can be regarded as the potential added on the ideal spring oscillator.
  Here, we set \(k = m = 1\) for simplicity.

\begin{figure}[h!]
  \captionsetup{justification=centering}
  \begin{minipage}[t]{0.5\textwidth}
    \centering
    \includegraphics[scale=0.38]{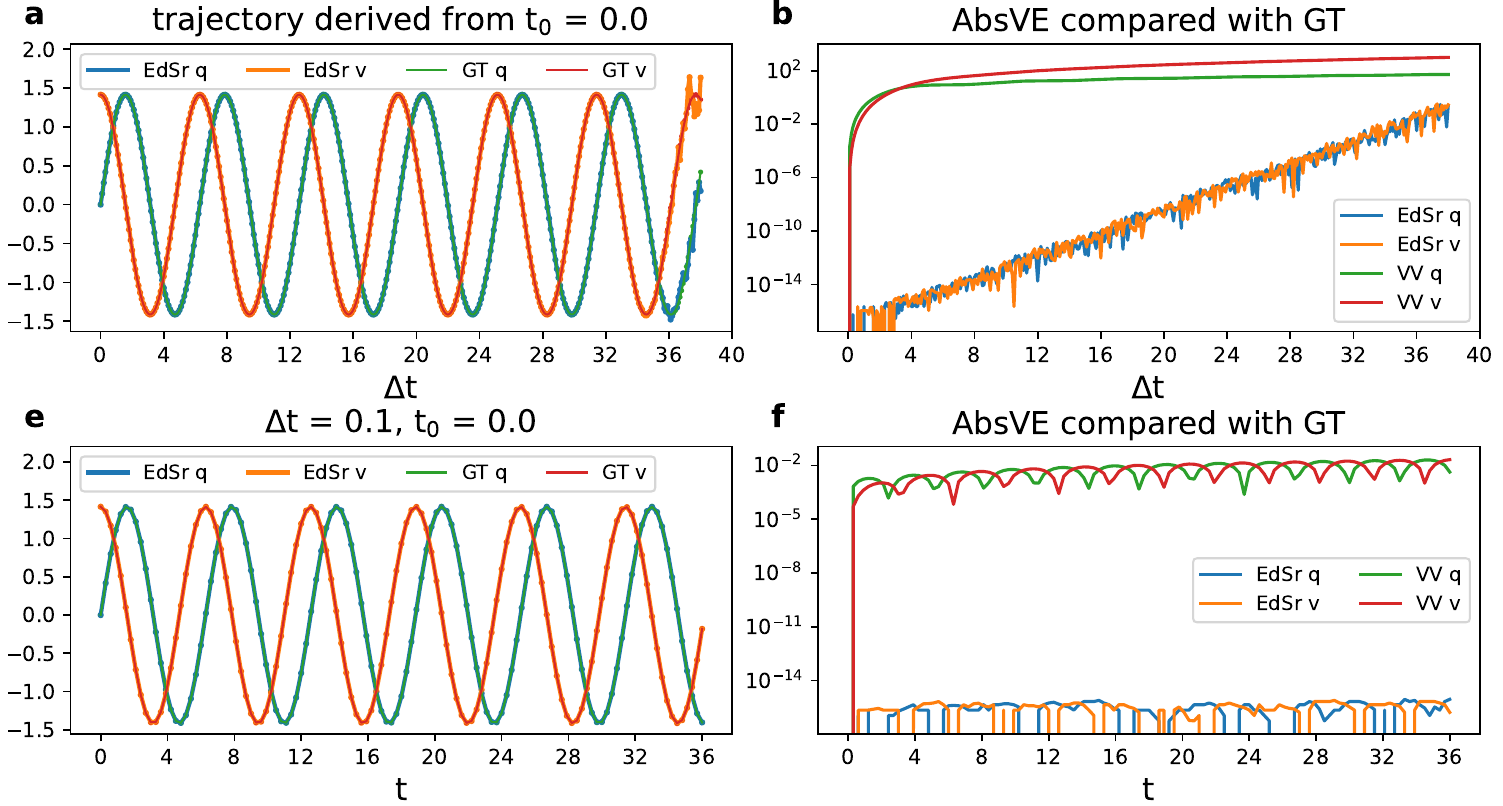}

  \end{minipage}
  \begin{minipage}[t]{0.5\linewidth}
    \centering
    \includegraphics[scale=0.38]{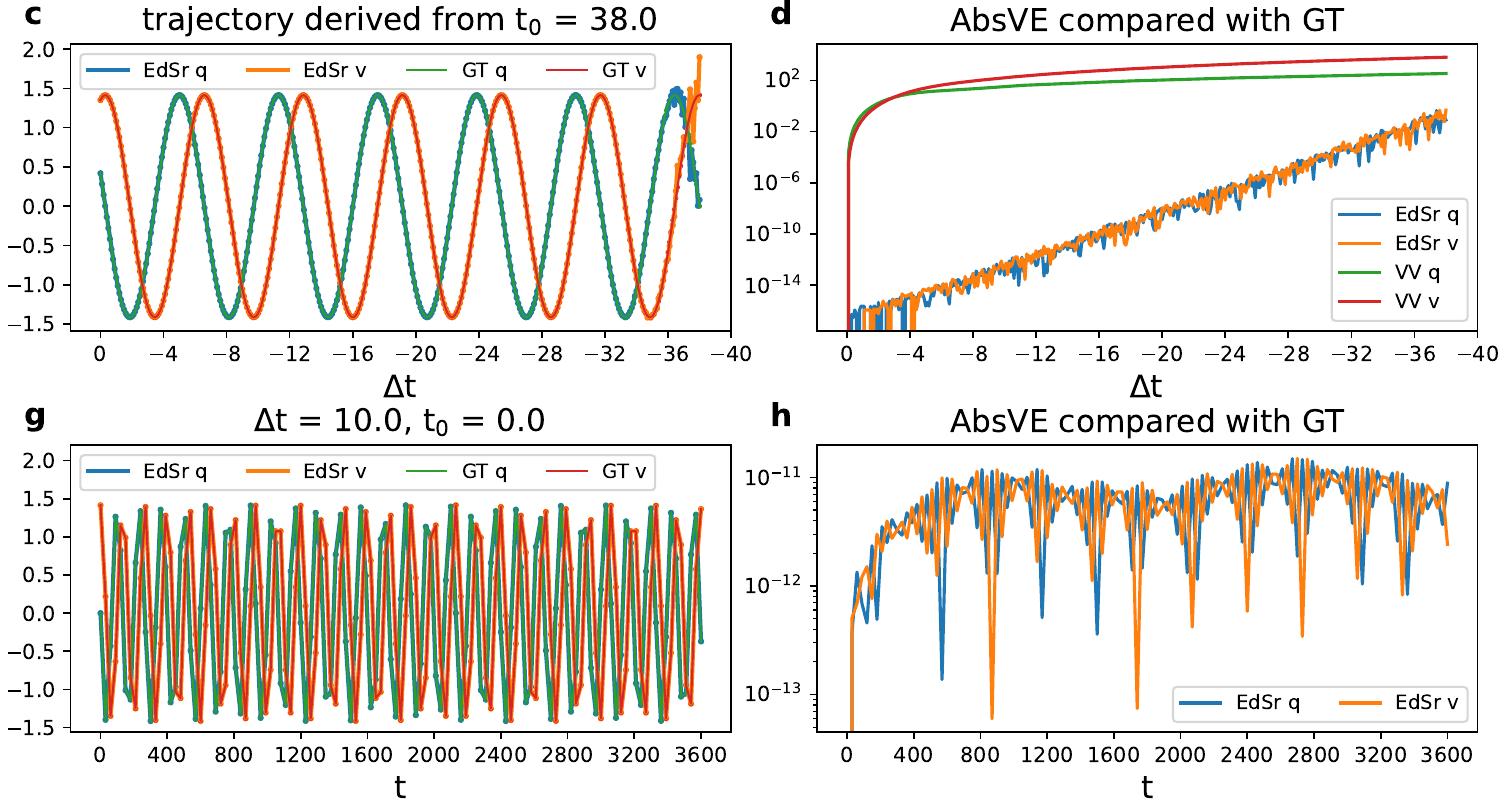}

  \end{minipage}
  \caption{
    Results of two groups of sub-experiments of ideal spring. 
    (a-d) show the results of the first group of sub-experiments, and (e-h) show the results of the second group of sub-experiments.
    (a) Blue line, orange line denote generalized coordinates \(q\), velocities \(v\) generated by EdSr with positive $\Delta t$, respectively.
       Green line, red line denote \(q\) and \(v\) obtained in GT, respectively. 
    (b) Blue line, orange line denote AbsVEs between EdSr and GT.
       Green line, red line denote AbsVEs between VV and GT. 
    (c), (d) are same as (a), (b),  but with negative \(\Delta t\).
    (e), (f) show the comparisons of $q$ and $v$ between EdSr and GT, and the corresponding AbsVEs with \(\Delta t = 0.1\).
    (g), (h) denote the results with \(\Delta t = 10.0\), the corresponding AbsVEs between VV and GT are not shown as they are very huge (over $1\times10^2$). 
  }
  \label{picture:spring}
\end{figure}

Similar to  the analysis of simple functions, 
 we conduct the similar two groups of sub-experiments on ideal spring (Figure \ref{picture:spring}).
  In the first group of sub-experiments,
 we show the variations of coordinates and the corresponding velocities over \(\Delta t\), as well as AbsVEs of EdSr and VV compared with GT (Figure \ref{picture:spring}a and b).
 We also perform the tests backwards with a negative timestep \(\Delta t\) (Figure \ref{picture:spring}c and d),   for the verification of time-reversal symmetry.
  The results show that EdSr achieves significantly higher precision than the VV integrator at same $\Delta t$ value by comparing with GT.
  The obvious deviation does not appear until \(\Delta t\) is greater than 35.0 for EdSr method. 
   Furthermore, EdSr is able to handle negative timestep \(\Delta t\), as evidenced by the agreement between the results of Figure \ref{picture:spring}a and c, as well as Figure \ref{picture:spring}b and d. It indicates that EdSr shows excellent performance for the systems with time symmetry.
  In the second group of sub-experiments, we generate the trajectories of ideal spring at $\Delta t=0.1$ and 10.0  (Figure \ref{picture:spring}e-h).
 Both the coordinates and velocities generated by EdSr remain very close to the GT results at both small $\Delta t=0.1$ and large $\Delta t=10.0$  timesteps (Figure \ref{picture:spring}e and g),
 and the AbsVEs are substantially smaller at the two selected $\Delta t$ values than those of VV integrator (Figure \ref{picture:spring}f and h).
   Additional results by setting $\Delta t=1.0$ and 5.0 are provided in Figure S5.
  Thus, 
  it is unquestionable that EdSr outperforms the  VV integrator for the ideal spring model.

\textbf{Ideal Pendulum}\label{ideal-pendulum} 

  The Hamiltonian of ideal pendulum without friction can be described as:
\begin{equation}
    \mathcal{H} = mgl(1 - \cos \theta) + \frac{1}{2} ml^2 v^2
\end{equation}
where $m$, $g$, $l$, \(\theta\) and \(v\) are the mass of pendulum, gravitational constant, length of pendulum, angle and angular velocity respectively.
 The first term gravity  serves as the external potential on the ideal pendulum. 
  To simplify the experiment, we set $m = l = 1$, and $g = 4$.
  The angle \(\theta\) and angular velocity \(v\) are initially set to \(\displaystyle \frac{\pi}{3}\) and 0, respectively.

\begin{figure}[h!]
  \captionsetup{justification=centering}
  \begin{minipage}[t]{0.5\textwidth}
    \centering
    \includegraphics[scale=0.38]{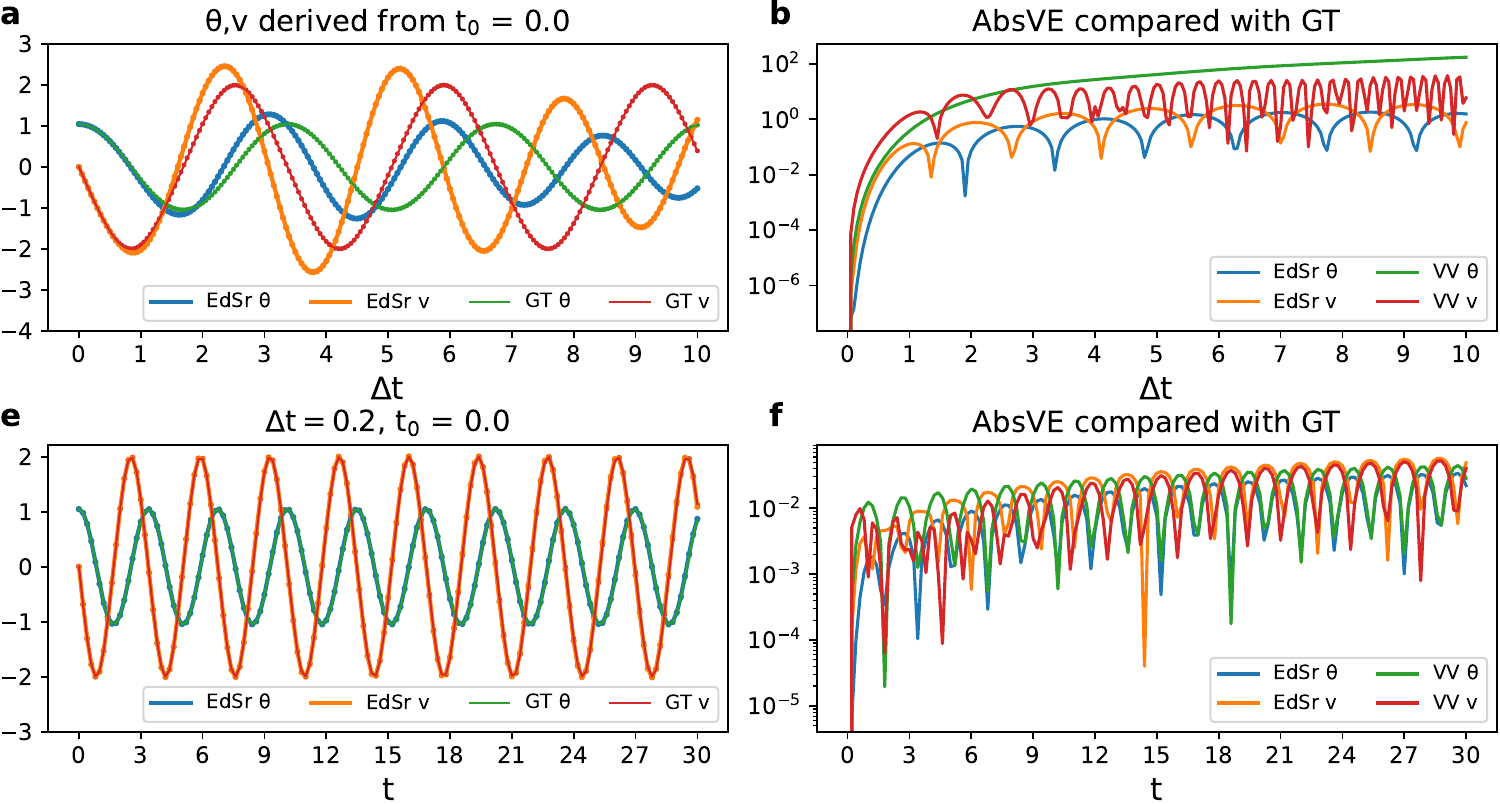}

  \end{minipage}
  \begin{minipage}[t]{0.5\textwidth}
    \centering
    \includegraphics[scale=0.38]{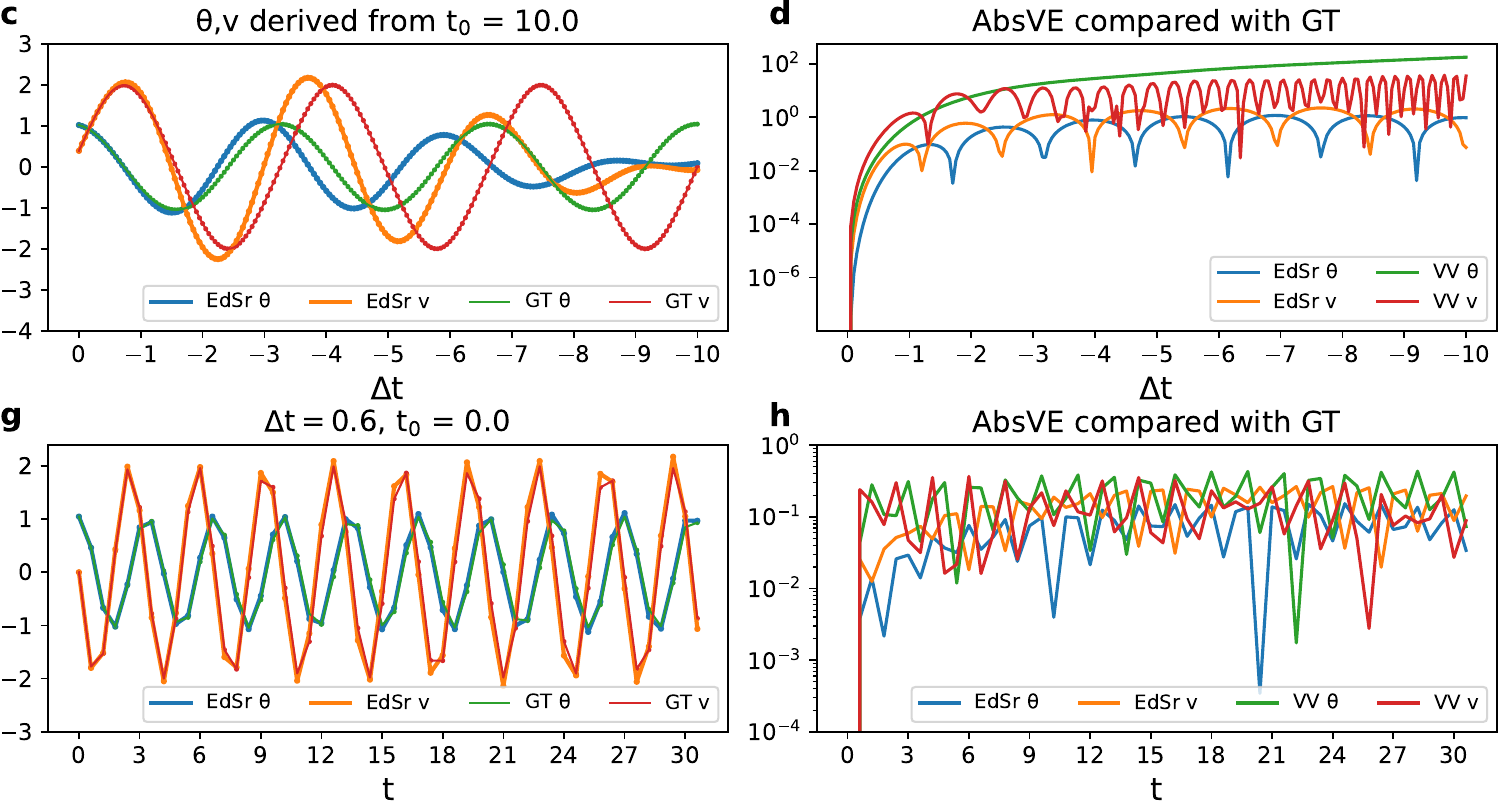}
  \end{minipage}
  \caption{
      Results of two groups of sub-experiments of ideal pendulum. 
    (a-d) show the results of the first group of sub-experiments, and (e-h) show the results of the second group of sub-experiments.
    (a) Blue line, orange line denote angles \(\theta\), angular velocities \(v\) generated by EdSr with positive $\Delta t$, respectively.
        Green line, red line denote angles \(\theta\), angular velocities \(v\) generated by GT, respectively. 
    (b) Blue line, orange line denote AbsVEs between EdSr and GT;
        green line, red line denote AbsVEs between VV and GT. 
    (c) and  (d) are same as (a) and (b), respectively, but with negative \(\Delta t\).
    (e), (f) show the comparisons of $\theta$ and $v$ between EdSr and GT, and the corresponding AbsVEs with \(\Delta t = 0.2\).
    (g), (h) denote the results with \(\Delta t = 0.6\). 
  }
  \label{picture:pendulum}
\end{figure}

In the first group of sub-experiments, 
  Figure \ref{picture:pendulum}a presents the results of angles \(\theta\) and angular velocities \(v\) for positive values of  \(\Delta t\)  obtained through EdSr method, along with the comparison with GT.
 The corresponding  reverse time evolutions (negative \(\Delta t\)) of \(\theta\) and \(v\) are exhibited in Figure \ref{picture:pendulum}c.
  The results of \(\theta\) and \(v\) obtained via EdSr are close with those from GT, although the deviations increase with large absolute $\Delta t$.
  We also investigate the AbsVEs of EdSr compared with GT, together with those from VV integrator (Figure \ref{picture:pendulum}b and d).
 Although AbsVE values increase with larger absolute \(\Delta t\) for both EdSr and VV,
 EdSr produces smaller deviations in \(\theta\) and \(v\) compared to the VV integrator.
  The results shown in Figure \ref{picture:pendulum}a-d also demonstrate that EdSr exhibits time symmetry.
  Figure \ref{picture:pendulum}e-h display the results of the second group of sub-experiments with $\Delta t=0.2$ and 0.6.
 EdSr shows smaller deviation than VV integrator in long time simulations with the selective \(\Delta t\) values.
  Additional results  for the ideal pendulum with $\Delta t=\pm 1.0$ are shown in Figure S6, where EdSr and VV exhibit comparable performance.
  
\textbf{Two-Body Model}\label{two-body-model} 

In the two-body model, 
 one particle interacts with the other particle by Law of universal gravitation.
  Therefore, the Hamiltonian for two-body system can be described as:
\begin{equation}
    \mathcal {H} = \frac{1}{2}(m_1 + m_2) V^2 + \frac{1}{2} \mu (v_1^2 + v_2^2) + G\frac{m_1m_2}{|q_1 - q_2|}
\end{equation}
where \(m_1\), \(q_1\), \(v_1\) are mass, coordinates, velocity of one particle, respectively;
  \(m_2\), \(q_2\), \(v_2\) are the counterparts.
 \(G\) is the universal gravitational constant, 
 \(\mu\) is the reduced mass and \(V\) is the velocity of center of mass.
  The last term represents the interaction between the two particles in the model. 
  For simplicity, we set \(G = m_1 = m_2 = 1\), \(V = 0\). 

\begin{figure}
  \centering
  \captionsetup{justification=centering}
  \includegraphics[scale=0.75]{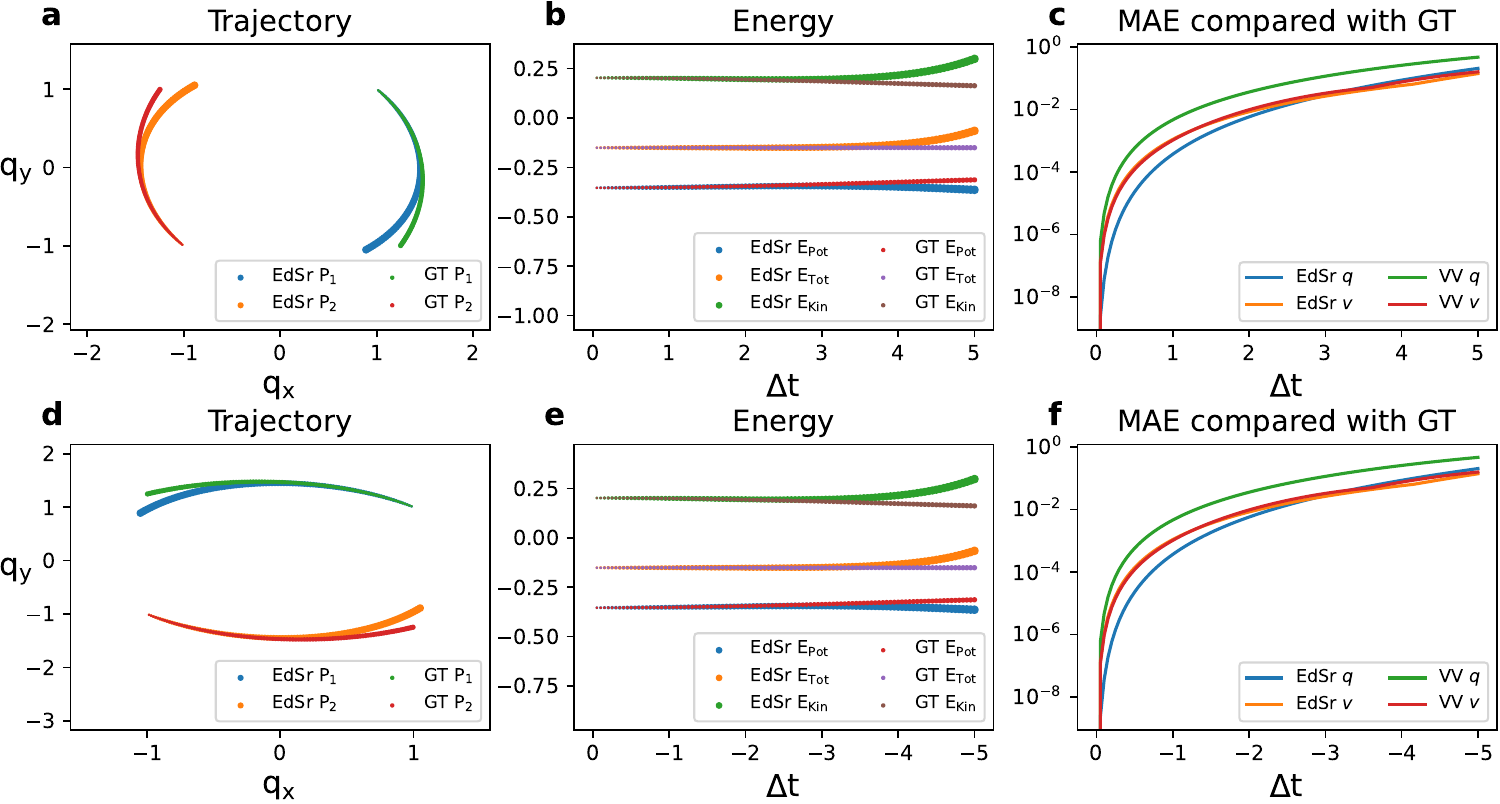}
  \caption{
    Results of the first group of sub-experiment of two-body model. 
    (a), (d) denote the positions of the particles in two-body (P$_1$ and P$_2$) generated by EdSr and GT, the symbol sizes increase as the absolute \(\Delta t\) values vary from 0.0 to 5.0. 
    (b), (e) exhibit kinetic energies (\(\mathrm{E_{Kin}}\), green line and brown line), 
    total energies (\(\mathrm{E_{Tot}}\), orange line and purple line),
    potential energies (\(\mathrm{E_{Pot}}\), red line and blue line) from EdSr and GT over \(\Delta t\).
    (c), (f) denote the MAEs between EdSr and GT (blue and orange lines), and compare with that between VV and GT (green and red lines) over \(\Delta t\).
  }
  \label{picture:two_body_init}
\end{figure}

In the first group of sub-experiments of two-body model, 
 we also firstly evaluate the performance of EdSr on both positive \(\Delta t\) and negative \(\Delta t\).
We visualize the trajectories of particles (Figure \ref{picture:two_body_init}a and d), and calculate thermodynamic properties including
 total energies (\(\mathrm{E_{Tot}}\)), kinetic energies (\(\mathrm{E_{Kin}}\)) and potential energies (\(\mathrm{E_{Pot}}\)) (Figure \ref{picture:two_body_init}b and e).
 The trajectories and energies obtained with EdSr are consistent with GT when $\Delta t<2.0$, afterwards the deviations become apparent for these properties.
  For the comparison of calculated position \(q\) and velocity \(v\) with VV integrator, we compute the Mean Absolute Error (MAE) between the results from EdSr and GT, as well as between VV and GT, respectively. 
  The MAE is defined as:

\begin{equation}
    \text{MAE} = \frac{1}{4}\sum_{i = 1}^{N:2}\sum_{j = 1}^{ndim:2}|\text{X}_{ij} - \text{GT}_{ij}| \label{eq:MAE}
\end{equation}

  where $N=2$ denotes the number of particles in the two-body model, $ndim=2$ indicates the two-dimensional nature of the system, X represents the outputs from either EdSr or VV group.
  The results of MAEs generated from EdSr and VV by comparing with GT are exhibited in Figure \ref{picture:two_body_init}c and f, at positive $\Delta t$ and negative $\Delta t$ values, respectively.
  The deviations of position \(q\) appear when the absolute \(\Delta t\) is greater than 2.0 for both EdSr and VV integrator. 
  However, the MAE of position \(q\) obtained with EdSr is significantly lower than that with VV integrator, indicating the superior accuracy of EdSr method.
The MAEs for velocity \(v\) are comparable between EdSr and the VV integrator.

\begin{figure}
  \centering
  \captionsetup{justification=centering}
  \includegraphics[scale=0.75]{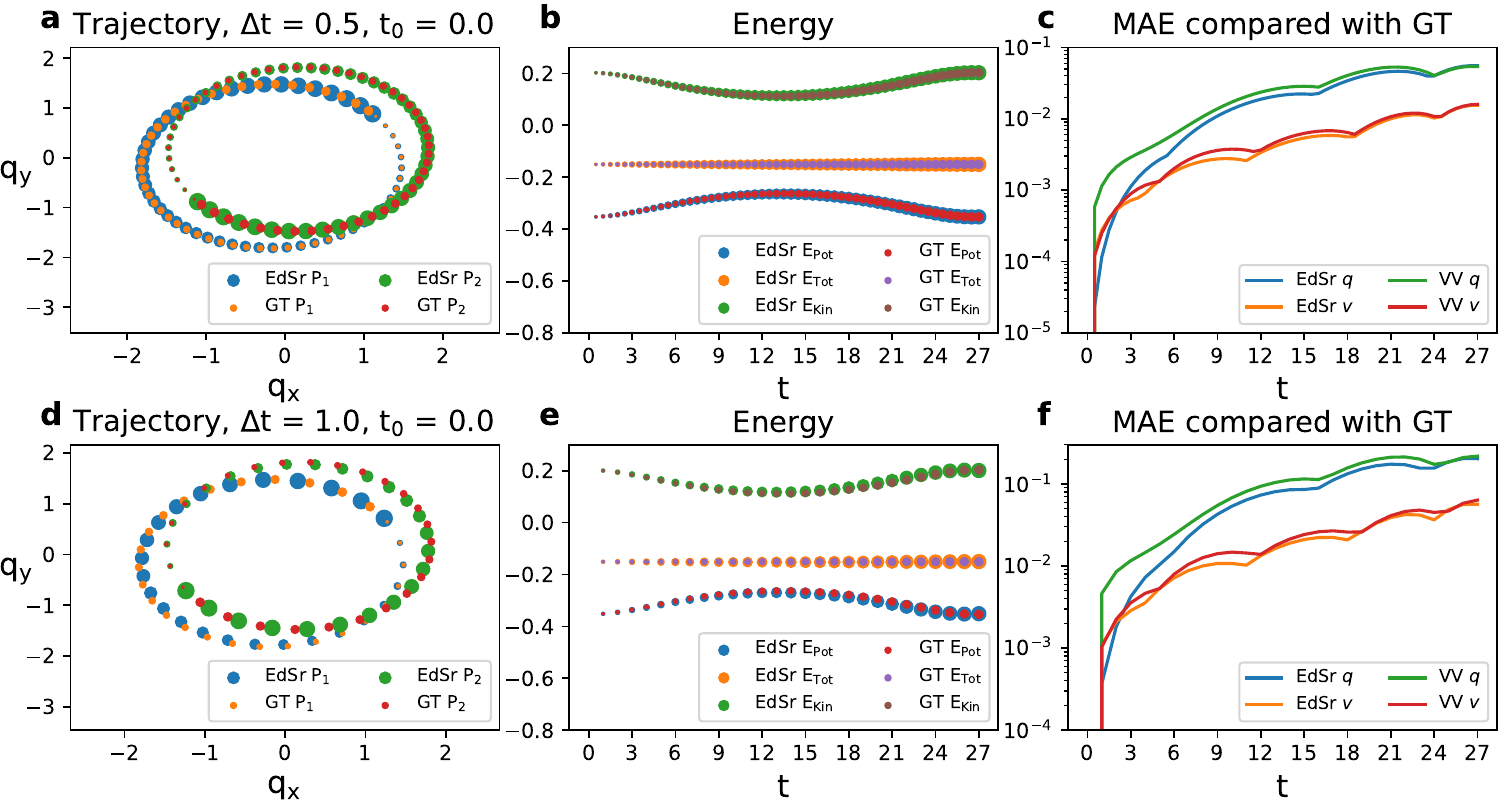}
  \caption{
    Results of the second group of sub-experiments of two-body model. 
    (a), (d) denote the time evolutions of positions of the particles in two-body (P$_1$ and P$_2$)  generated by EdSr and GT, the symbol sizes increase as the time starts from 0.0 to 27.0.
    The time evolutions of different energies obtained by EdSr and GT are exhibited in (b), (e).
    (c), (f) denote the MAEs between EdSr and GT, and compare with those between VV and GT over simulation time.
  }
  \label{picture:two_body_cur_0.5_1.0}
\end{figure}

In the second group of sub-experiments of two-body model, 
 we generate only one period of data for the predefined $\Delta t$ values, in order to enable strict comparisons of EdSr with GT and VV.
  Here we mainly show the results of \(\Delta t = 0.5\) and \(\Delta t = 1.0\) in Figure \ref{picture:two_body_cur_0.5_1.0}.
  The  trajectories of two particles obtained with EdSr are similar as those from GT (Figure \ref{picture:two_body_cur_0.5_1.0}a and d), except that ``a forward effect'' appears in EdSr, 
 that is, it appears slightly ``faster'' than GT.
 ``The forward effect'' becomes more pronounced as the increase of  \(\Delta t\).

The time evolutions of potential energy, total energy and kinetic energy of EdSr closely match those of GT at both $\Delta t=0.5$ and 1.0 (Figure \ref{picture:two_body_cur_0.5_1.0}b and e).
  EdSr demonstrates higher accuracy in $q$ and $v$ than VV integrator in short-time simulation (\(t < 21.0\)), through the MAE results compared with GT (Figure \ref{picture:two_body_cur_0.5_1.0}c and f).
  But the performance of EdSr becomes comparable to that of VV integrator if the simulation time \(t > 21.0\).
We also examine the results generated by EdSr at negative $\Delta t$ with identical absolute values (Figure S7), which show the same tendency as Figure \ref{picture:two_body_cur_0.5_1.0}.
The results of second group of sub-experiments of two-body model with other $\Delta t$ values are shown in Figure S8.

\subsection{All-Atom MD Simulation}\label{all-atom-molecular-dynamics-simulation}

Here, in order to evaluate the application of EdSr method in MD simulation,
we employ a previously studied system involving the diffusion of indole in beta zeolite, namely \textit{indole@Beta zeolite}, as the all-atom MD simulation test system \cite{Zeolite}. The simulations are carried out using the LAMMPS software.
The zeolite structure is optimized using a core-shell force field \cite{core-shell-force-field_1,core-shell-force-field_2,core-shell-force-field_3}, while the interface force field \cite{interface-force-field_1,interface-force-field_2} is adopted for the MD simulations of the zeolite.
The OPLS force field is used for indole molecule. 
The detailed system information can be found in our earlier publication \cite{Zeolite}.
 Firstly, we perform the simulation in the canonical ensemble (NVT) for 1.0 ns, with the temperature at 700 K controlled by Nos\'{e}-Hoover thermostat\cite{Nose-nvt,Hoover-nvt}, 
 in order to ensure the system reaches thermodynamic equilibrium.
 Afterwards, we conduct three groups of simulations in the micro-canonical ensemble (NVE):
  the benchmark group (\textbf{BM group}) using VV integrator with a small timestep \(\Delta t\) = 0.01 fs;
  the control group (\textbf{MD group}) using VV integrator with larger timesteps \(\Delta t\) = 1.0, 2.0, 3.0 fs,
 $\Delta t=3.0$ fs is the largest timestep we can perform the simulation for the system with VV integrator;
  \textbf{EdSr group} using EdSr method with the timesteps \(\Delta t\) = 1.0, 2.0, 3.0, 4.0 fs.
 The MD and EdSr groups are conducted for \(1\times 10^4\) steps, and the BM group is performed for the same simulation time as MD and EdSr groups.
  So far, as we have not yet embedded EdSr into LAMMPS software, 
 we only take LAMMPS software as ``force engine'' and use Python script to run EdSr, which could increase the time cost for performing simulation with EdSr potentially.
  However, unlike the EdSr, we use LAMMPS and their build-in ``run'' API to run simulations for BM and MD groups.
  Here, we also calculate MAE for characterizing the performance of EdSr and MD groups, which is redefined as:

\begin{equation}
  \text{MAE}\ =\  
  \left\{
  \begin{aligned}
  & \frac{1}{N*M} \sum_{i = 1}^{N:natoms} \sum_{j = 1}^{M:ndims} |\text{X}_{ij} - \text{BM}_{ij}| &  \qquad (a) \\ \label{eq:MAE_MD}
  & \frac{1}{N} \sum_{i = 1}^{N:time} |\text{X}_{i} - \text{BM}_{i}| &  \qquad (b)
  \end{aligned}
  \right. 
\end{equation}
where X denotes the outputs from either EdSr or MD group. 
  We use Equation \ref{eq:MAE_MD}a to compute MAE of three dimensional ($ndims=3$) coordinates, while Equation \ref{eq:MAE_MD}b to compute MAEs of some scalar quantities such as energies, 
root mean square deviations (RMSD), and so on.

\begin{figure}
  \centering
  \captionsetup{justification=centering}
  \includegraphics[scale = 0.75]{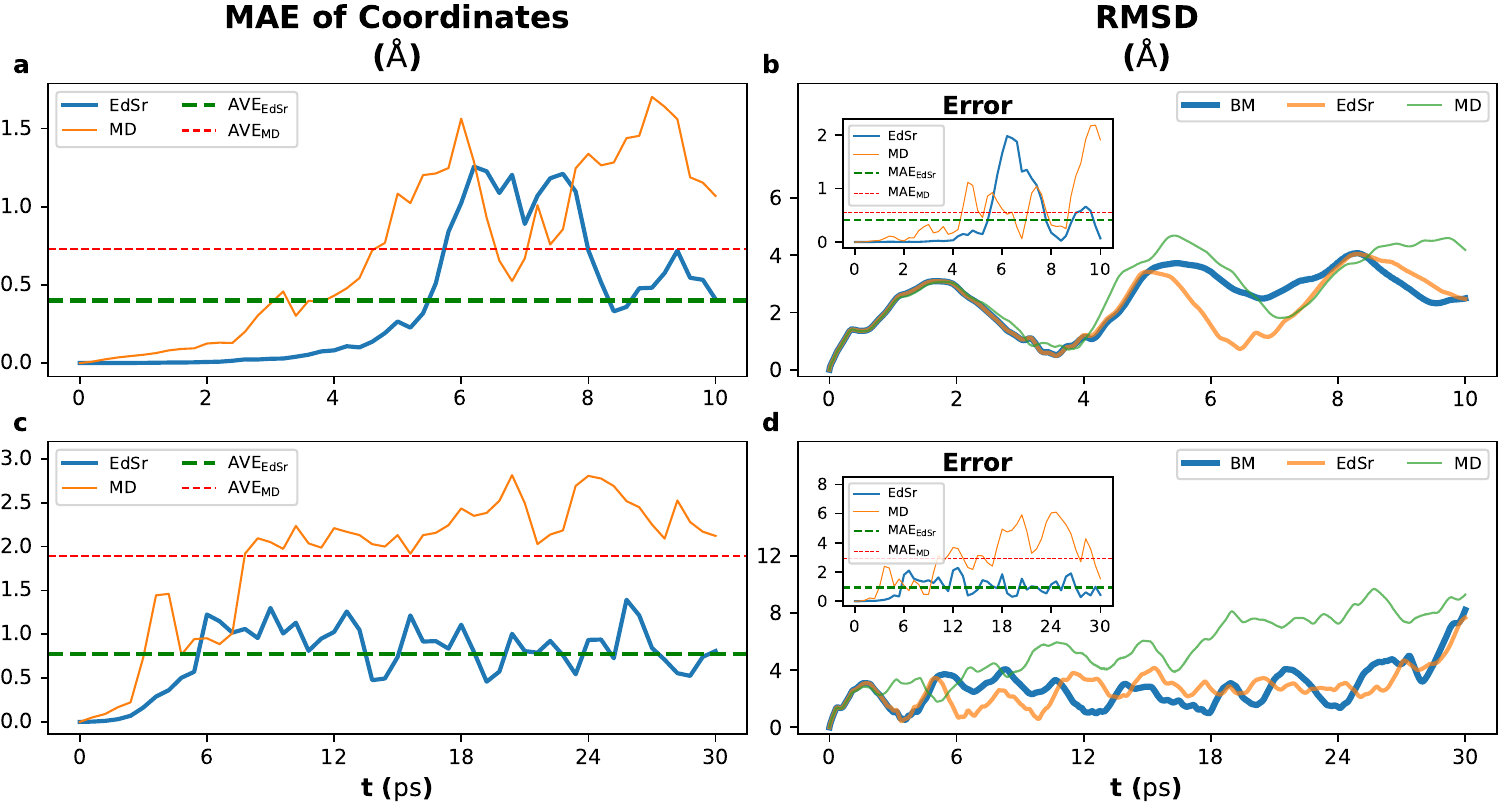}
  \caption{
     MAEs between EdSr and BM groups, as well as between MD and BM groups  of the atom coordinates of an indole molecule over time at $\Delta t=1.0$ fs (a) and 3.0 fs (c), the two dashed lines in the two sub-figures represent the average MAEs.
    Time evolutions of RMSDs of BM, EdSr and MD groups over time at  $\Delta t=1.0$ fs (b) and 3.0 fs (d), the insets are the AbsVEs between EdSr and BM, as well as between MD and BM groups, with their MAEs shown as the dashed lines.
  }
  \label{picture:indole_coord_rmsd_1.0_3.0}
\end{figure}

  First, we  select one representative indole molecule in the system randomly, and  compare the MAEs of the indole molecule coordinates between EdSr and BM groups, as well as between MD and BM groups.
  The MAE between EdSr and BM is smaller than that between MD and BM when $t<6.0$ ps at $\Delta t=1.0$ fs (Figure \ref{picture:indole_coord_rmsd_1.0_3.0}a), then the two methods generate similar results at longer times.
  The MAE from EdSr method is significantly smaller than that from MD at $\Delta t=3.0$ fs (Figure \ref{picture:indole_coord_rmsd_1.0_3.0}c), demonstrating the higher accuracy of EdSr in generating trajectories at larger timestep. This observation is further supported by the averaged MAEs from the comparisons between EdSr and MD (dashed lines in Figure \ref{picture:indole_coord_rmsd_1.0_3.0}a and c).
 Thus, 
 EdSr outperforms MD group on constructing trajectories of indole molecules at various timesteps.
The RMSDs from BM, MD and EdSr groups at $\Delta t=1.0$ and 3.0 fs are shown in Figure \ref{picture:indole_coord_rmsd_1.0_3.0}b and d, respectively.
The RMSD results obtained via MD and EdSr groups are similar at the smaller timestep of 1.0 fs, 
and EdSr exhibits better performance than MD at the larger timestep of 3.0 fs.

\begin{figure}
  \centering
  \captionsetup{justification=centering}
  \includegraphics[scale=0.75]{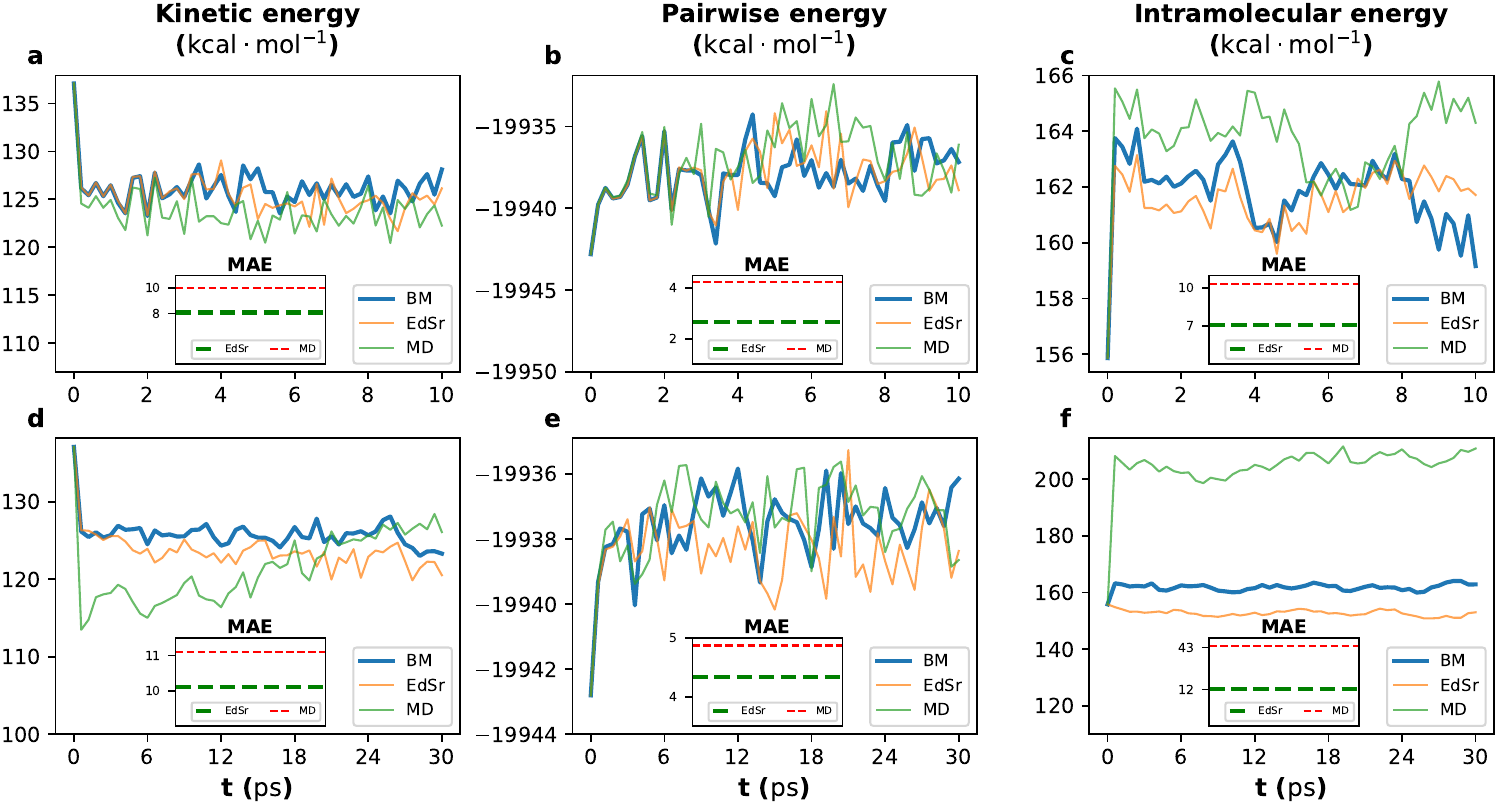}
  \caption{
    Time evolutions of different energies obtained by BM, EdSr and MD groups in the all-atom MD simulations. (a-c) show the results of \(\Delta t\) = 1.0 fs,
    (d-f) show the results of \(\Delta t\) = 3.0 fs.
  } 
  \label{picture:indole_ke_epair_emol_press_1.0_3.0}
\end{figure}

In order to  evaluate the performance of EdSr more comprehensively, 
we compare thermodynamic properties by computing the time evolutions of total kinetic energy, total non-bonded pairwise energy, and intramolecular energy of indoles (Figure \ref{picture:indole_ke_epair_emol_press_1.0_3.0}),
 and then we also compute the MAEs of different energies between EdSr and BM groups, as well as MD and BM groups (insets in Figure \ref{picture:indole_ke_epair_emol_press_1.0_3.0}).
  Generally speaking, 
 the EdSr method shows smaller deviation than VV integrator across all thermodynamic properties at both \(\Delta t\) = 1.0 fs (Figure \ref{picture:indole_ke_epair_emol_press_1.0_3.0}a-c) and \(\Delta t\) = 3.0 fs (Figure \ref{picture:indole_ke_epair_emol_press_1.0_3.0}d-f) for the system.
  Especially at \(\Delta t\) = 3.0 fs,
 the MAE of intramolecular energy obtained via EdSr keeps at a significantly lower level than MD group throughout the entire simulation (Figure \ref{picture:indole_ke_epair_emol_press_1.0_3.0}f).
   We calculate the radial distribution functions (RDF) between indoles and zeolites, in which both EdSr and MD results match the BM group (Figure S9a and c),
 indicating similar relative positions of indole molecules across different simulations. 
  However, we find that the deviations of structural and thermodynamical properties between EdSr and BM increase over time (Figure \ref{picture:indole_coord_rmsd_1.0_3.0} and \ref{picture:indole_ke_epair_emol_press_1.0_3.0}), 
 leading to slight discrepancies in velocity distribution of indoles on both EdSr and MD group after finishing the whole simulations (Figure S9b and d).
 The comparisons of EdSr and MD groups versus BM at $\Delta t=2.0$ fs, as well as the results of EdSr and BM groups at $\Delta t=4.0$ fs are provided from Figure S10 to Figure S12,
 which show that the structural properties obtained via EdSr are consistent with BM group at $\Delta t=4.0$ fs, although the deviations of energies emerge.

\subsection{Coarse-Grained MD Simulation}\label{coarse-grained-molecular-dynamics-simulation}

In order to explore the performance of EdSr in more complicated systems with larger time and length scales,
 we set up a protein aqueous solution system based on ubiquitin in CG level \cite{ubiquitin}.
 The Martini coarse-grained (CG) force field version 2.2 is used to convert the atomistic protein structure into the corresponding CG representation \cite{martinize}. Each simulation box contains a single ubiquitin molecule represented by 163 CG beads, solvated in water CG beads which include 2224 P$_4$ beads and 247 BP$_4$ beads to prevent the solvent from freezing\cite{Martini-beads-type}.
  First, 
 we perform a short energy minimization followed by an equilibration of the CG system for 6 ns. 
  Afterwards, we use the equilibrated CG configuration as the initial structure of the following three groups:
  a benchmark group (\textbf{BM group}) using VV integrator with a  small timestep \(\Delta t\) = 1.0 fs; 
  a control group  (\textbf{MD group}) using VV integrator with large timesteps \(\Delta t\) = 10.0 fs, 20.0 fs;
  an \textbf{EdSr group} using EdSr method with  timesteps \(\Delta t\) = 10.0 fs, 20.0 fs, 30.0 fs.
 The MD and EdSr simulations are also conducted under NVE ensemble for \(1\times10^4\) steps, and the BM group performs for the same simulation time as the other two groups.

\begin{figure}[h]
  \centering
  \captionsetup{justification=centering}
  \includegraphics[scale=1.0]{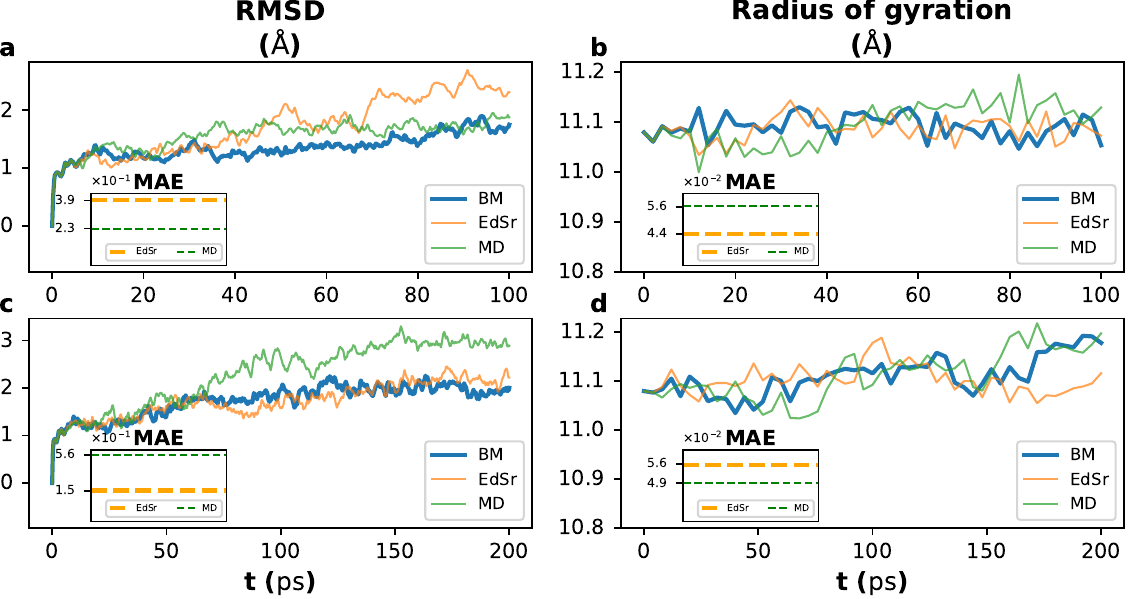}
  \caption{    
    Time evolutions of RMSDs of BM, EdSr and MD groups at  $\Delta t=10.0$ fs (a) and 20.0 fs (c), the insets are the MAEs of RMSDs from EdSr and MD groups by comparing with BM group.
    (b) and (d) shows radii of gyration of ubiquitin molecules in the three groups over time and their MAEs  at  $\Delta t=10.0$ fs  and 20.0 fs, respectively.
  }
  \label{picture:ubiquitin_rmsd_rg_10.0_20.0}
\end{figure}

We compute the RMSDs from the three groups at $\Delta t=10.0$ and 20.0 fs, which are shown in Figure \ref{picture:ubiquitin_rmsd_rg_10.0_20.0}a and c, respectively.
The MAE of RMSD between EdSr and BM groups is slightly larger than that between MD and groups at $\Delta t=10.0$ fs (inset of Figure \ref{picture:ubiquitin_rmsd_rg_10.0_20.0}a), but becomes significantly smaller at $\Delta t=20.0$ fs (inset of Figure \ref{picture:ubiquitin_rmsd_rg_10.0_20.0}c). 
In addition, we investigate the configuration of ubiquitin by calculating the radii of gyration in the three groups  at $\Delta t=10.0$ and 20.0 fs, the results are exhibited in Figure \ref{picture:ubiquitin_rmsd_rg_10.0_20.0}b and d, respectively.
Both the results from EdSr and MD groups are close to the BM group at the two selected timesteps, which indicates that the EdSr method can reproduce the conformation of macromolecule accurately.

\begin{figure}
  \centering
  \captionsetup{justification=centering}
  \includegraphics[scale=0.95]{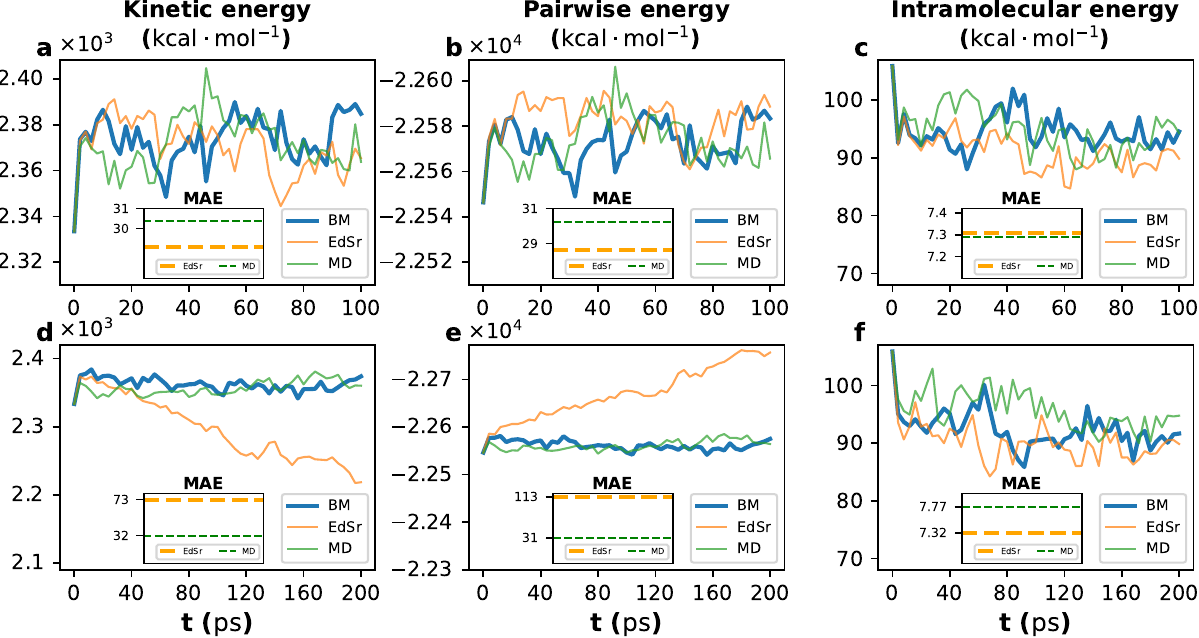}
  \caption{
    Kinetic energy (a,d), pairwise energy (b,e), 
    intramolecular energy (c,f) of BM, EdSr and MD  groups over time.
    (a-c) shows the results of \(\Delta t\) = 10.0 fs and (d-f) shows the results of \(\Delta t\) = 20.0 fs.
    The insets exhibit the MAEs of EdSr and MD by comparing with BM group.
  }
  \label{picture:ubiquitin_ke_epair_emol_press_10.0_20.0}
\end{figure}

Then, we also compare thermodynamics properties including kinetic energy, pairwise energy, intramolecular energy in the three groups at $\Delta t=10.0$ and 20.0 fs, the results are presented in Figure \ref{picture:ubiquitin_ke_epair_emol_press_10.0_20.0}.
 The kinetic, pairwise and intromolecular energies obtained via EdSr are similar as those from BM and MD groups at $\Delta t=10.0$ fs (Figure \ref{picture:ubiquitin_ke_epair_emol_press_10.0_20.0}a-c).
  However, the kinetic and pairwise energies from EdSr appear to be non-conservative at $\Delta t=20.0$ fs (Figure \ref{picture:ubiquitin_ke_epair_emol_press_10.0_20.0}d and e), which results in larger MAE of EdSr by comparing with BM group (insets of Figure \ref{picture:ubiquitin_ke_epair_emol_press_10.0_20.0}d and e).
  In contrast, the MAE of intramolecular energy from the EdSr group remains lower than that from the MD group,
 which indicates that the conformation of protein keeps stable in the EdSr group at large timestep (Figure \ref{picture:ubiquitin_ke_epair_emol_press_10.0_20.0}f).
  We compare the radial distribution functions between ubiquitin and water beads across the three groups.
  Despite the deviations of kinetic and pairwise energies appear in EdSr at $\Delta t=20.0$ fs, 
 the distributions of water beads around the ubiquitin  are very similar in all three groups (Figure S13a and c), 
 as well as the velocity distributions of the CG beads (Figure S13b and d).
 Furthermore, we also perform the EdSr group at $\Delta t=30.0$ fs and compare the results with the BM group (the MD group can not run at such a large timestep). The structural properties, thermodynamic features, radial distribution functions and velocity distributions are shown in Figure S14, which also confirm that the EdSr group produce consistent  RMSD, radius of gyration, and radial distribution function results compared with the BM group.
  In order to investigate the effect of system complexity on the accuracy of EdSr, 
 we design another set of simulations with reduced degrees of freedom by removing water beads, and evaluate the performance of EdSr at $\Delta t=20.0$ and 30.0 fs, which are shown from Figure S15 to S17.
 The differences of kinetic energies and potential energies between EdSr and BM groups are significantly smaller in the simplified systems than the systems with water CG beads.
This suggests that the deviation of energies from EdSr may be related to the number of degrees of freedom for a specific system,
  it suggests that systems with more degree of freedom might exhibit larger deviation using EdSr.

\section{Conclusions}\label{conclusion}

This paper proposes EdSr, 
 a computational method based on Taylor expansion and ordinary differential equation, 
 to extend the time scale of MD simulation flexibly and dynamically.
  Unlike machine learning based method that require extensive training, 
  EdSr endeavors to  generate next state correctly from a specific state without cost of extra training.
  We not only design a series of experiments, including simple functions, 
 ideal physical models and MD simulations, to verify the feasibility of EdSr,
 but also evaluate the performance of EdSr by comparing with VV integrator. 
 Overall, the results demonstrate that
 EdSr exhibits remarkable capability  to perform at larger integration timestep in some systems  (e.g. ideal spring, all-atom model of \textit{indole@Beta zeolite}), while it does not perform very well in some other cases (like $f(x) = x^3$ function, coarse-grained model of protein aqueous solution). 
It seems that the performance of EdSr depends on the specific problem.

The development of EdSr method is motivated by the following reasons:
 First, we transform the Taylor expansion into a kind of recursive process,
 which makes it possible to use the recursive process to approximate Taylor Expansion with infinite term,
 and can eliminate the requirement for setting multiple parameter sets manually when selecting different partial differential equations or models.
  Second, we provide a special perspective that establishes a connection between Taylor expansion and Markov process. 
  Mathematically, we demonstrate how EdSr can be transformed into a second-order Markov chain,
 which probably makes us understand the working mode of EdSr using probability density function. 
  More importantly, EdSr makes it possible to scale up the \(\Delta x\) in Taylor series, 
 enabling tasks to be handled through a parallel computing framework. 
  For example, if EdSr can be applied at \(\Delta t = 10\) for a given system, 
 we can first  generate the states at \(t = 0, 1, ..., 9\) in parallel,
 and then use these states to compute the subsequent states at \(t = 10, 11, ..., 19\), also in parallel.

However, there is still more room for the improvement of EdSr approach. 
  First, 
it seems that EdSr does not perform equally well across all systems, because some functions or models might exceed the hypotheses outlined in \textbf{Methods}.
Therefore, it is essential to evaluate EdSr before embedding it into a specific system.
  Then, the maximum timestep of EdSr depends on the system it is adopted.
In addition, executing a single step of EdSr is still time consuming, 
and the maximum iteration time $N$ seems to be related to the specific system in which EdSr is applied. A detailed discussion about the setting of maximum iteration is provided in Supporting Information Section 10.
  In future work, 
  we will focus on addressing these issues and strive to reduce the computational cost of EdSr.
  
  Recently, some studies suggest that numerical difference method has been applied to machine learning 
  such as enhancing convolutional neural networks \cite{cnn-highorder-NDM}, Neural ODE\cite{chen_neural_2018}, Flow Matching,
which also creates opportunities for integrating such methods with EdSr.  Besides, we are going to explore the applications of EdSr in areas 
 such as chemical reaction dynamics, normalizing flows, etc.
  Although EdSr still has some limitations, we believe that it will be a promising approach in the future.

\section*{Code and data availability}\label{code-availability}

All codes about this paper are available at \url{https://github.com/haiming-Cao/EdSr}.
\section*{Supporting Information}
The Supporting Information is available free of charge, including the theory on which EdSr method is based,  derivation for displacement and velocity of EdSr, discussion of the maximum iteration time $N$, and additional figures.

\section*{Acknowledgement}\label{Acknowledgement}
We appreciate the financial support from the Guangdong Basic and Applied Basic Research Foundation (2023A1515012535, 2022A1515010873) and National Natural Science Foundation of China (no. 22003080). 

\section*{Author Declarations}
\subsection*{Conflict of interest}
The authors have no conflicts to disclose.

\providecommand{\latin}[1]{#1}
\makeatletter
\providecommand{\doi}
  {\begingroup\let\do\@makeother\dospecials
  \catcode`\{=1 \catcode`\}=2 \doi@aux}
\providecommand{\doi@aux}[1]{\endgroup\texttt{#1}}
\makeatother
\providecommand*\mcitethebibliography{\thebibliography}
\csname @ifundefined\endcsname{endmcitethebibliography}
  {\let\endmcitethebibliography\endthebibliography}{}

\clearpage

\section*{Table of Contents}

  \centering
  \includegraphics[width=8.3cm]{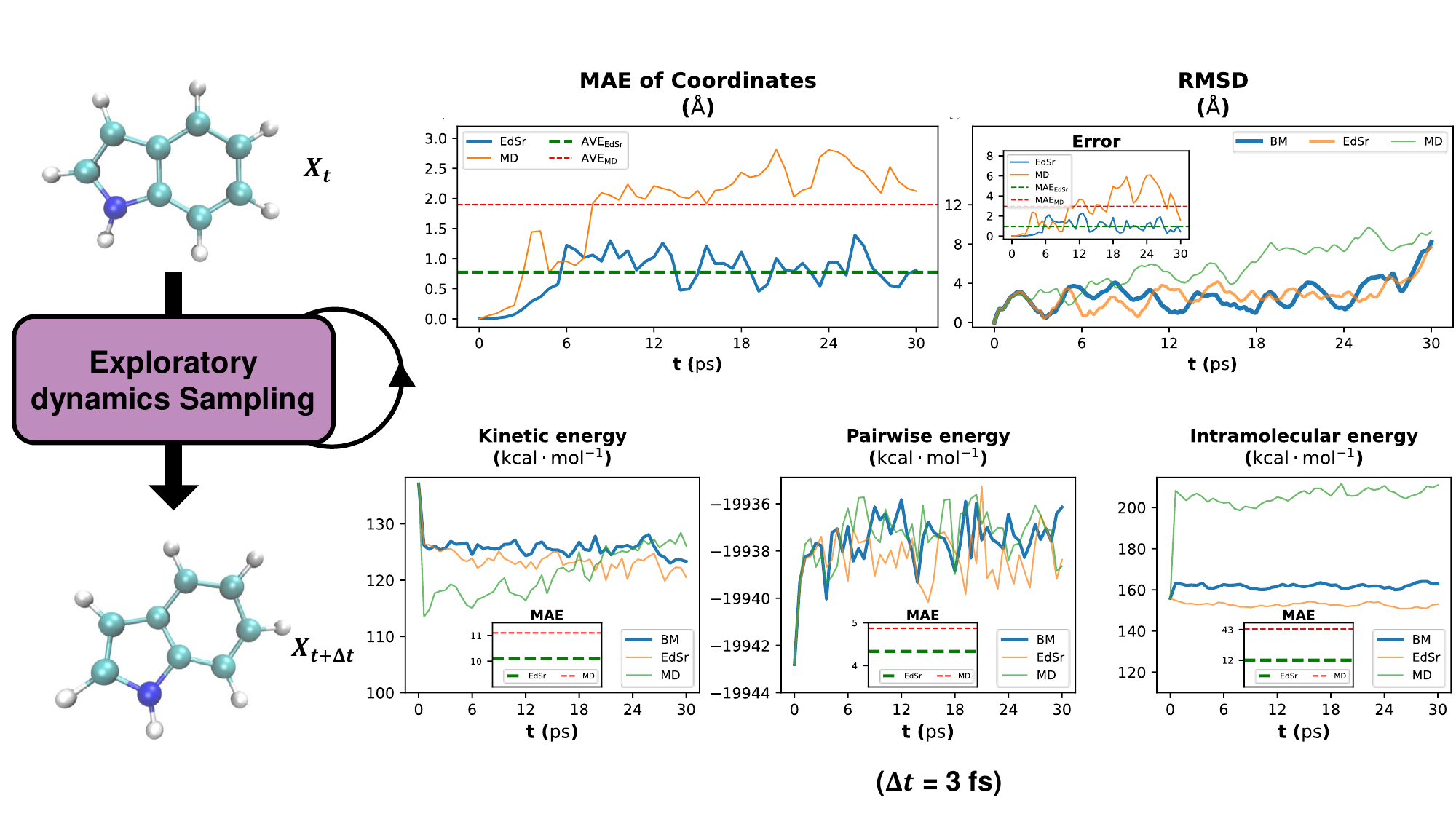}

\end{document}


\date{}
\maketitle
\tableofcontents

\freefootnote{$^1$School of Chemical Engineering and Technology, Sun Yat-Sen University, Zhuhai 519082, China.
E-mail: libin76@mail.sysu.edu.cn}

\newpage

\section{Derivation for Displacement and Velocity of EdSr}\label{introduction}
\subsection{Derivation for the Part of Displacement of EdSr}
Here, we use three examples ($N=2$, $N=4$, $N=6$) to illustrate the execution process of EdSr.
First, we recall Equation (5)  in main article here, with ``the second term'' omitted:
\begin{align} \label{eq:terms}
  X(b) \approx & X_N + DX_N\Delta t + \frac{1}{2!} D^2X_N\Delta t^2 + ... + \frac{1}{N!} D^NX_N\Delta t^N
\end{align}
where \(D\), \(X_N\) denote the differential operator, $X(a)$, respectively. 
If \(N = 2\), Equation \ref{eq:terms} can be written as, 
\begin{equation}
  \begin{aligned}
    X(b) &\approx X_2 + DX_2\Delta t + \frac{1}{2} D^2\underline{X_2}\Delta t^2 \\
    &\approx \frac{1}{0!}D^0 \bigg(X_2 + \frac{1}{1!} (DX_2\Delta t + \frac{1}{2} D^2X_2\Delta t^2)\bigg)\Delta t^0 \\
    &\approx \underline{\mathbf{X_0}}
  \end{aligned}
\end{equation}
which is equivalent to solving equation of motion. 
We assume that the result of equation of motion is \(\mathbf{X_0}\).

If \(N = 4\), Equation \ref{eq:terms} becomes,
\begin{equation}\label{eq:terms_4}
  X(b) \approx X_4 + DX_4\Delta t + \frac{1}{2} D^2X_4\Delta t^2 + \frac{1}{3!} D^3X_4\Delta t^3 + \frac{1}{4!} D^4X_4\Delta t^4
\end{equation}
we first merge the last three terms in Equation \ref{eq:terms_4}. After getting the intermediate value \(\mathbf{X_2}\), we  use the equation of motion to get \(\mathbf{X_0}\).
\begin{equation}
  \begin{aligned}
    X(b) &\approx X_4 + DX_4\Delta t + \frac{1}{2!} D^2X_4\Delta t^2 + \frac{1}{3!} D^3X_4\Delta t^3 + \frac{1}{4!} D^4\underline{X_4}\Delta t^4 \\
    &\approx X_4 + DX_4\Delta t + \frac{1}{2!} D^{2}\bigg(X_4 + \frac{1}{3}\Big(DX_4\Delta t + \frac{1}{4}D^2X_4\Delta t^{2}\Big)\bigg)\Delta t^{2} \\
    &\approx \frac{1}{0!}D^0 X_4 \Delta t^0 + \frac{1}{1!}D^1 X_4 \Delta t^1 + \frac{1}{2} D^2\underline{\mathbf{X_2}}\Delta t^2, \qquad \mathrm{where}\ \mathbf{X_2} = X_4 + \frac{1}{3}DX_4\Delta t + \frac{1}{4}D^2X_4\Delta t^{2} \\
    &\approx \frac{1}{0!}D^0 \bigg(X_4 + \frac{1}{1!} (DX_4\Delta t + \frac{1}{2} D^2\mathbf{X_2}\Delta t^2)\bigg)\Delta t^0 \\
    &\approx \underline{\mathbf{X_0}}
  \end{aligned}
\end{equation}
If \(N = 6\), we can make the following derivation:
\begin{equation}
  \begin{aligned}
    X(b) &\approx X_6 + DX_6\Delta t + \frac{1}{2!} D^2X_6\Delta t^2 + \frac{1}{3!} D^3X_6\Delta t^3 + \frac{1}{4!} D^4X_6\Delta t^4 + \frac{1}{5!} D^5X_6\Delta t^5 + \frac{1}{6!} D^6\underline{X_6}\Delta t^6 \\
    &\approx X_6 + DX_6\Delta t + \frac{1}{2!} D^2X_6\Delta t^2 + \frac{1}{3!} D^3X_6\Delta t^3 + \frac{1}{4!} D^{4}\bigg(X_6 + \frac{1}{5}\Big(DX_6\Delta t + \frac{1}{6}D^2X_6\Delta t^{2}\Big)\bigg)\Delta t^{4} \\
    &\approx X_6 + DX_6\Delta t + \frac{1}{2!} D^2X_6\Delta t^2 + \frac{1}{3!} D^3X_6\Delta t^3 + \frac{1}{4!} D^{4}\underline{\mathbf{X_4}}\Delta t^{4}, \quad \mathrm{where}\ \mathbf{X_4} = X_6 + \frac{1}{5}\Big(DX_6\Delta t + \frac{1}{6}D^2X_6\Delta t^{2}\Big) \\
    &\approx X_6 + DX_6\Delta t + \frac{1}{2!} D^{2}\bigg(X_6 + \frac{1}{3}\Big(DX_6\Delta t + \frac{1}{4}D^2\mathbf{X_4}\Delta t^{2}\Big)\bigg)\Delta t^{2} \\
    &\approx X_6 + DX_6\Delta t + \frac{1}{2} D^2\underline{\mathbf{X_2}}\Delta t^2, \qquad \qquad \mathrm{where}\ \mathbf{X_2} = X_6 + \frac{1}{3}DX_6\Delta t + \frac{1}{4}D^2\mathbf{X_4}\Delta t^{2} \\
    &\approx \underline{\mathbf{X_0}}
  \end{aligned}
\end{equation}

\subsection{Derivation for the Part of Velocity of EdSr}
Here, we present a derivation for the part of velocity of EdSr, starting from the definition of derivative,
\begin{equation}
  X'(t+\Delta t) ={} \lim_{a \to 0} \frac{X(t+ \Delta t + a) - X(t + \Delta t)}{a} \label{eq:definition_derivative}
\end{equation}
where $X(t)$ and \(\Delta t\) are given.
We substitute \(X(t+ \Delta t + a)\) and \(X(t + \Delta t)\) using Equation (5) in main article. Thus, Equation (\ref{eq:definition_derivative}) can be rewritten as:
\begin{align}
X'(t+\Delta t) ={} & \lim_{a \to 0} \frac{\bigg[X(t)+\sum\limits^{N}\limits_{n=1} \frac{1}{n!}D^{n}X(t)(\Delta t+a)^n\bigg]-\bigg[X(t)+\sum\limits_{n=1}^{N} \frac{1}{n!}D^{n}X(t)\Delta t^n\bigg]}{a} \notag \\
X'(t+\Delta t) ={} & \lim_{a \to 0} \frac{\sum\limits^{N}_{n=1} \frac{1}{n!}D^{n}X(t)(\Delta t+a)^n-\sum\limits_{n=1}^{N} \frac{1}{n!}D^{n}X(t)\Delta t^n}{a} \notag \\
X'(t+\Delta t) ={} & \lim_{a \to 0} \frac{\sum\limits^{N}_{n=1} \frac{1}{n!}D^{n}X(t)\bigg[(\Delta t+a)^n-\Delta t^n\bigg]}{a} \label{eq:result_derivative_edsr}
\end{align}
According to Binomial theorem, we subtract the terms in the square brackets in Equation (\ref{eq:result_derivative_edsr}), and then the above equation is changed as:
\begin{align}
X'(t+\Delta t) ={} & \lim_{a \to 0} \frac{a \cdot \sum\limits^{N}_{n=1}\frac{1}{n!}D^{n}X(t)\Delta t^{n-1}\cdot C_{n}^{1} + O(a)}{a} \notag\\
 ={} & \sum\limits^{N}_{n=1}\frac{1}{(n-1)!}D^{n}X(t)\Delta t^{n-1} \notag\\
={} & DX(t) + D^2X(t)\Delta t + \frac{1}{2!} D^3X(t)\Delta t^2 + \cdots + \frac{1}{(n-1)!} D^{n}X(t)\Delta t^{n-1} \notag\\
={} & \underbrace{DX(t) + D^2\bigg[\underbrace{X(t) + \frac{1}{2!}DX(t)\Delta t + \cdots + \frac{1}{(n-1)!}D^{n-2}X(t)\Delta t^{n-2}}_{\rm{\mathbf{inner}}}\bigg]\Delta t}_{\rm{\mathbf{outer}}}
\label{eq:inter_outer}
\end{align}

The ``inner'' form 
can be obtained by using the part of displacement in Equation (13) in main article, which 
corresponds to the first half in Equation (14) in main article. 
The derivation between them is similar as that for the displacement part of EdSr, which also starts from the last three terms in the ``inner'' form. 

\newpage
\section{Bridge between EdSr and Markov process}\label{bridge-between-edsr-and-markov-process}

In this section, we establish a connection between EdSr and Markov process. 
  Mathematically, we can transform EdSr into a second-order Markov chain.
  Here we  outline the transformation briefly.
In general, the initial velocities of particles in a system are sampled from Maxwell-Boltzmann distribution, 
which means \(v \sim \mathcal{N}(0, \frac{k_\mathrm{B}T}{M})\).
The logarithmic of probability density function \(\log p(x)\) can be described as the negative potential energy \(U(x)\) divided by \(k_\mathrm{B}T\)
(\(\log p(x) = - \frac{U(x)}{k_\mathrm{B}T}\)). 
By setting \(\frac{k_\mathrm{B}T}{M} = \sigma^2\), we can rewrite Equation (13) in main article as follows:

\begin{align}
  X_{n - 1} & ={} X_{N} + \Big[\frac{1}{2n - 1} - \frac{1}{2n}\Big] \sigma^2 \nabla_X \log p(X_n) \Delta t^{2} + \frac{1}{2n - 1} \sigma \Delta t \epsilon, \qquad \mathrm{where} \ \epsilon \sim \mathcal{N}(0, I) \notag \\
  X_{n - 2} & ={} X_{N} + \Big[\frac{1}{2n - 3} - \frac{1}{2n - 2}\Big] \sigma^2 \nabla_X \log p(X_{n - 1}) \Delta t^{2} + \frac{1}{2n - 3} \sigma \Delta t \epsilon, \qquad \mathrm{where} \ \epsilon \sim \mathcal{N}(0, I) \notag 
  \end{align}
Similarly, we can get the second-order Markov process by using \(X_{n-2} - X_{n - 1}\).

\begin{align*}
  X_{n - 2} - X_{n - 1} = {} & \sigma^2 \Delta t^2 \Bigg\{\Big[\frac{1}{2n - 3} - \frac{1}{2n - 2}\Big]\nabla_X \log p(x_{n - 1}) - \Big[\frac{1}{2n - 1} - \frac{1}{2n}\Big] \nabla_X \log p(x_{n})\Bigg\} \\
  & + \sigma \Delta t \Big(\frac{1}{2n - 3} - \frac{1}{2n - 1}\Big)  \epsilon, \qquad \mathrm{where} \ \epsilon \sim \mathcal{N}(0, I)
\end{align*}

\clearpage

\section{Simple function}\label{simple-function}

\subsection{Experimental Details}\label{experimental-details-simple-function}
Here, we take the \(f(x) = \sin x\) and \(f(x) = e^{0.1x}\) as examples.
The procedure for other functions is similar to that for sine and exponential functions.
In the experiments of simple functions, a key aspect is determining the values of the second-order derivatives..
In the simple function experiments, we  use the second-order derivatives directly, rather than  $-\frac{\nabla_XU(X)}{M}$ in Equation (13).
For the sine and exponential functions, the derivatives are calculated as follows:
\begin{align*}
  f(x)_{sin} ={} \sin x, \qquad & f(x)_{exp} ={} e^{0.1x} \\
  f'(x)_{sin} ={} \cos x, \qquad & f'(x)_{exp} ={} 0.1e^{0.1x} \\
  f''(x)_{sin} ={} -\sin x, \qquad & f''(x)_{exp} ={}0.01e^{0.1x} 
\end{align*}
Like potential energy function, if the equation is iterated successfully, the second-order derivative needs to be expressed as a function of $f(x)$.
We can rewrite their second-order derivatives by substituting the variables:
\begin{equation*}
  f''(x)_{sin} = -f(x), \qquad f''(x)_{exp} = 0.01f(x)
\end{equation*}
Once the second-order derivative is expressed properly, we can compute the tuple \((f(x+\Delta x), f'(x+\Delta x))\) for different \(\Delta x\).
In summary, if  the algorithm was implemented in ``pure'' function, a differential equation has to be constructed in the form of \(F(f''(x), f(x)) = 0\) for an explicit function.

\subsection{Results}\label{results-simple-function}

Here we show the results from EdSr for the functions \(y = \frac{1}{1 + e^{-x}}\) which is named sigmoid function, 
 as well as \(y = x^3\) and \(y = x^2 - 2x - 5\). 
  We use the source functions and their corresponding derivatives as ground truth (GT).
  

\begin{figure}[!ht]
  \centering
  \includegraphics[width=0.8\linewidth]{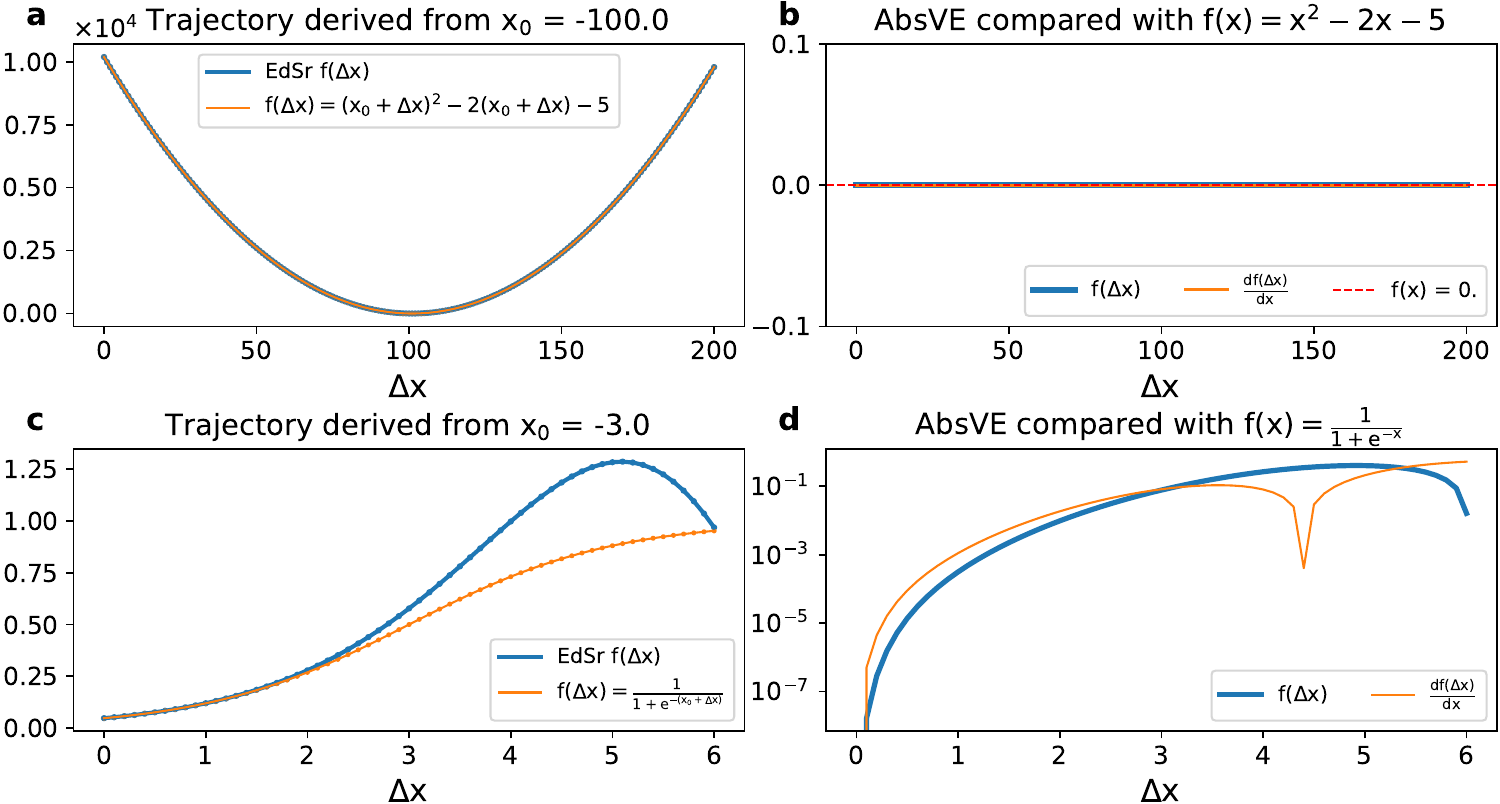}
  \captionsetup{justification=centering}
  \caption{
    Results of the first group of sub-experiments of \(f(x)= x^2 - 2x - 5\) and sigmoid function. 
    (a) GT of \(f(x) = x^2 - 2x - 5\) and the result generated by EdSr,
    (b) absolute value of error (AbsVE) between function \(f(x) = x^2 - 2x - 5\) and the corresponding derivative between GT and EdSr,
    (c) GT of \(f(x) = \frac{1}{1 + e^{-x}}\) and the result generated by EdSr,
    (d) AbsVE between function \(f(x) = \frac{1}{1 + e^{-x}}\) and the corresponding derivative between GT and EdSr.
  }
  \label{picture:x2-2x-5_sigmoid}
\end{figure}

\begin{figure}[!ht]
  \centering
  \includegraphics[width=0.8\linewidth]{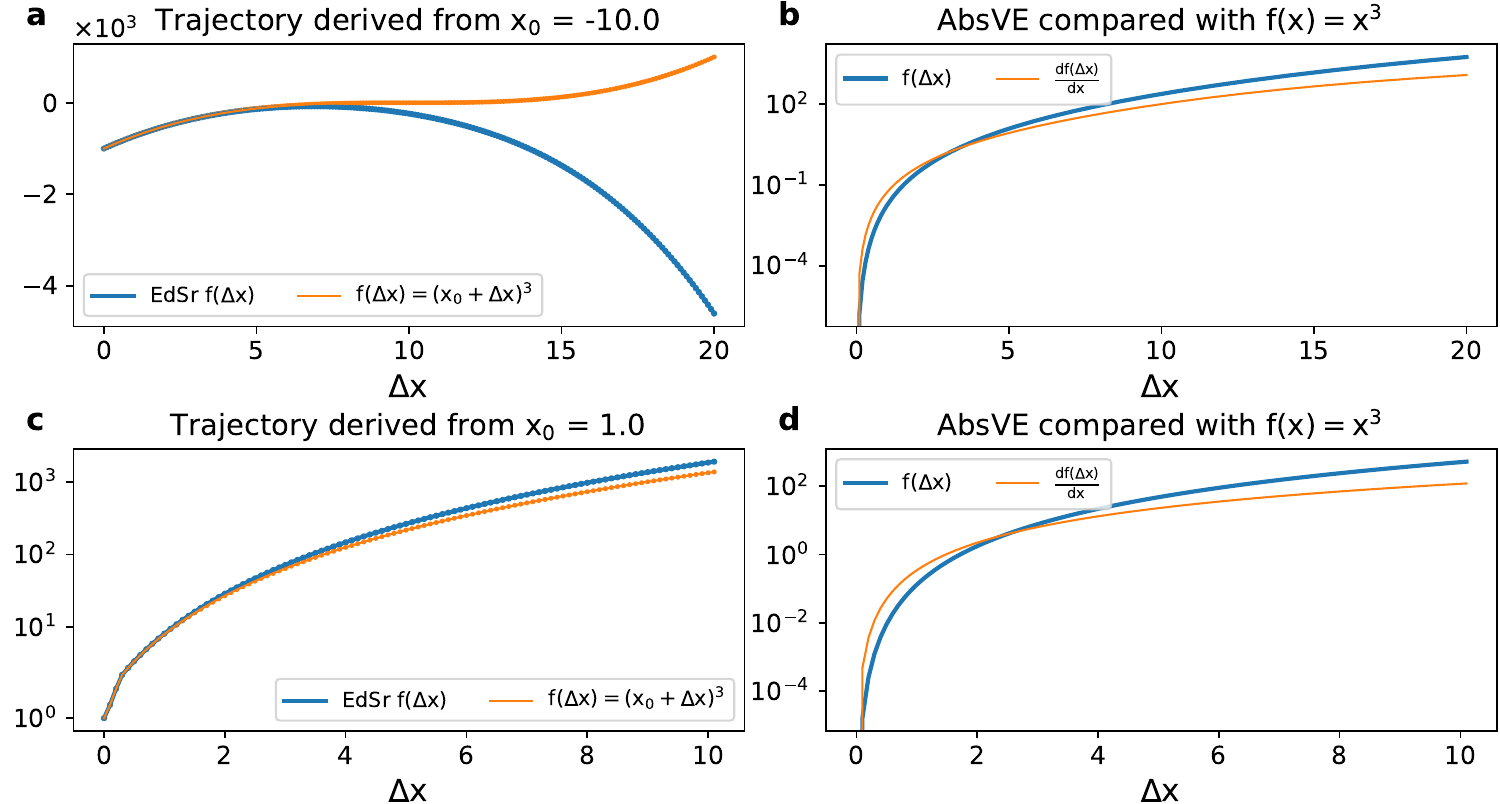}
  \captionsetup{justification=centering}
  \caption{
    Results of the first group of sub-experiments of \(f(x) = x^3\) with different initial point \(x_0\).
    (a), (c) denotes GT and the results generated by EdSr derived from $x_0=-10.0$ and 1.0, respectively.
    (b), (d) denotes the corresponding AbsVEs between GT and EdSr of the function and their derivatives.
  }
  \label{picture:x3_init}
\end{figure}


\begin{figure}[!ht]
  \centering
  \captionsetup{justification=centering}
  \includegraphics[width=0.9\linewidth]{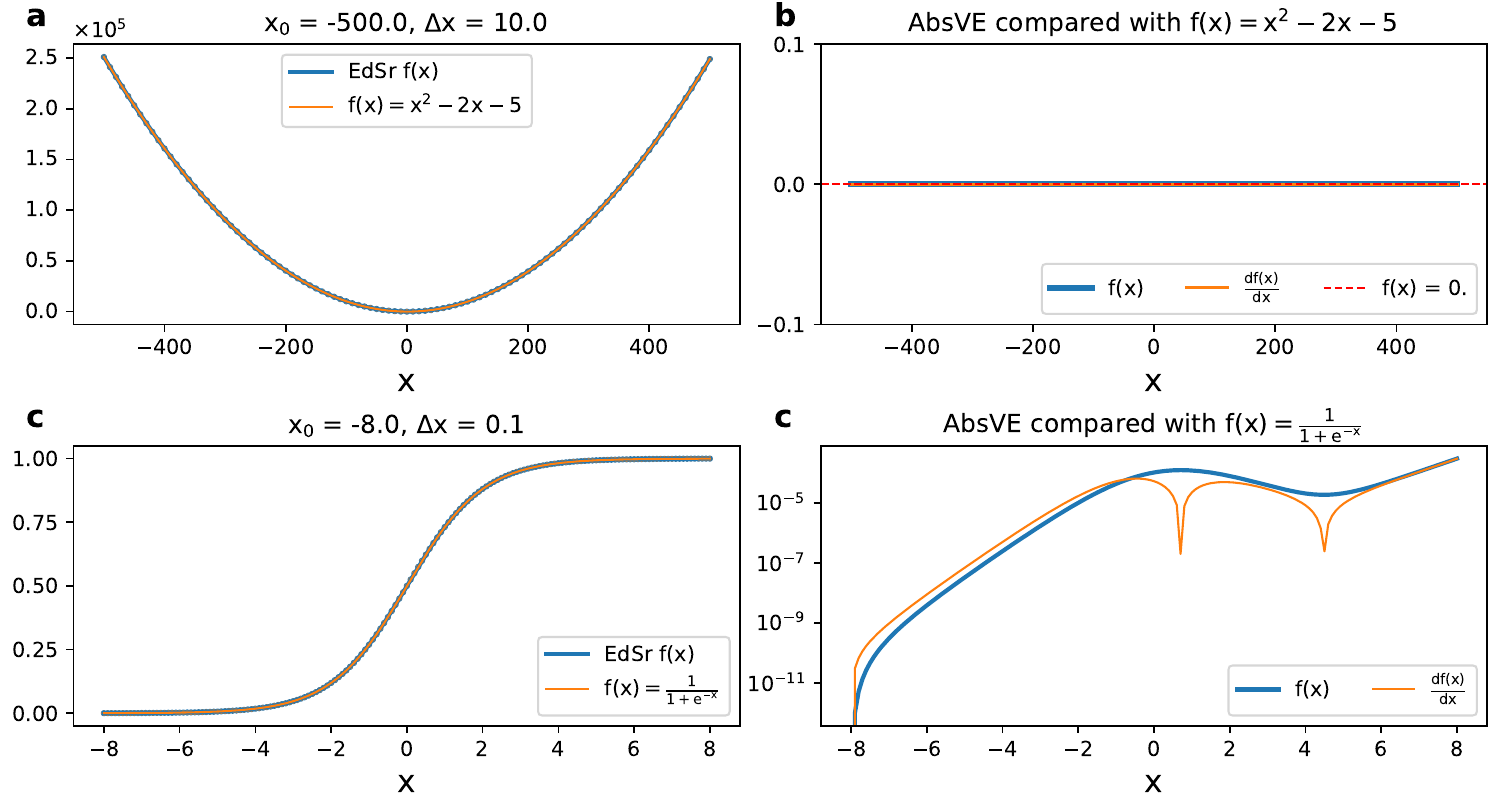}
  \caption{
    Results of the second group of sub-experiments of \(y = x^2 - 2x - 5\) and sigmoid function with different  initial point \(x_0\) and different \(\Delta x\).
  }
\end{figure}

\begin{figure}[!ht]
  \centering
  \captionsetup{justification=centering}
  \includegraphics[width=0.9\linewidth]{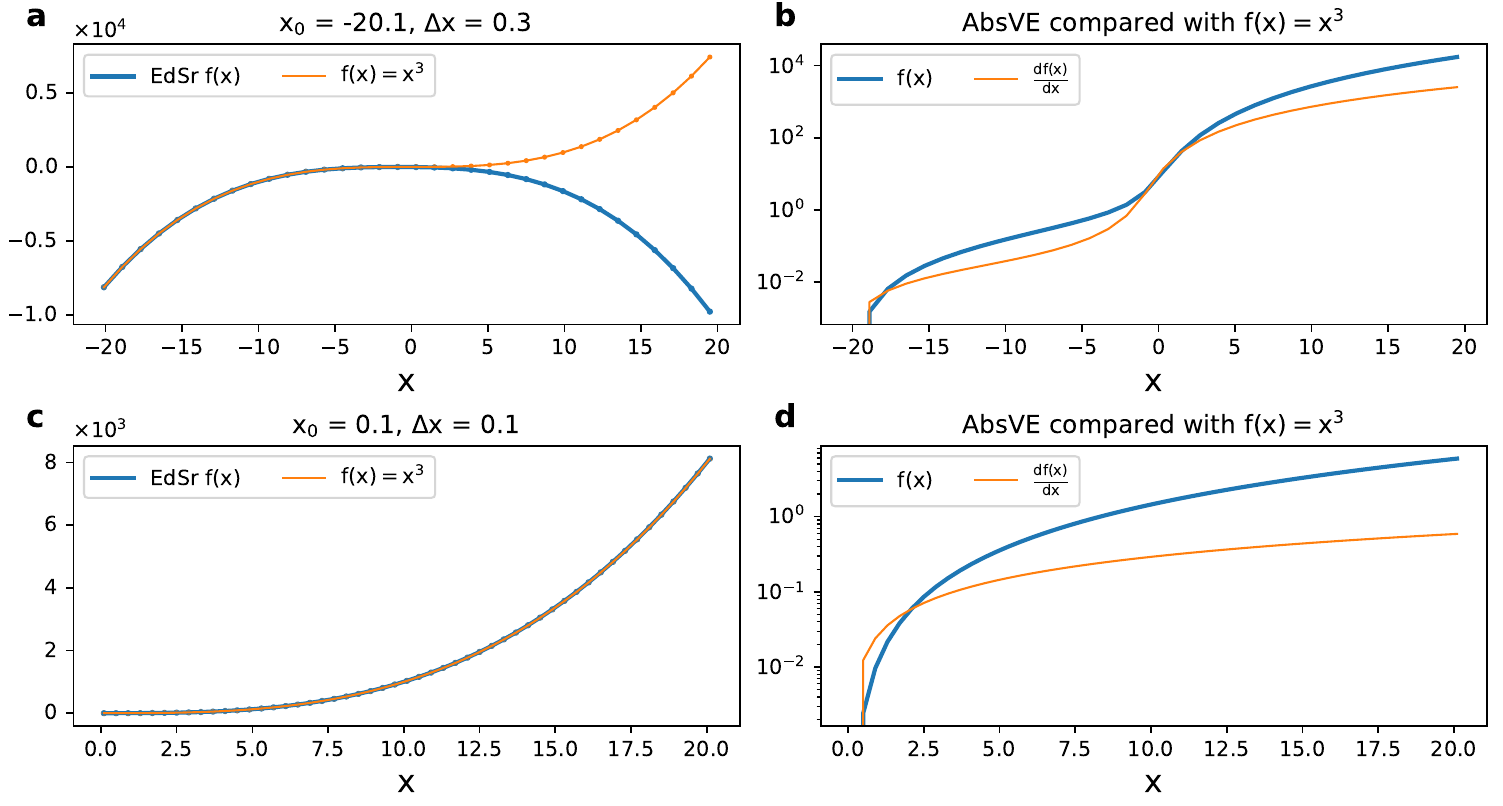}
  \caption{
    Results of the second group of sub-experiments of \(y = x^3\) with different initial point \(x_0\) and different \(\Delta x\).
  }
\end{figure}

\clearpage

\section{Ideal Spring}\label{ideal-spring}

\begin{figure}[!ht]
  \centering
  \captionsetup{justification=centering}
  \includegraphics[width=0.75\linewidth]{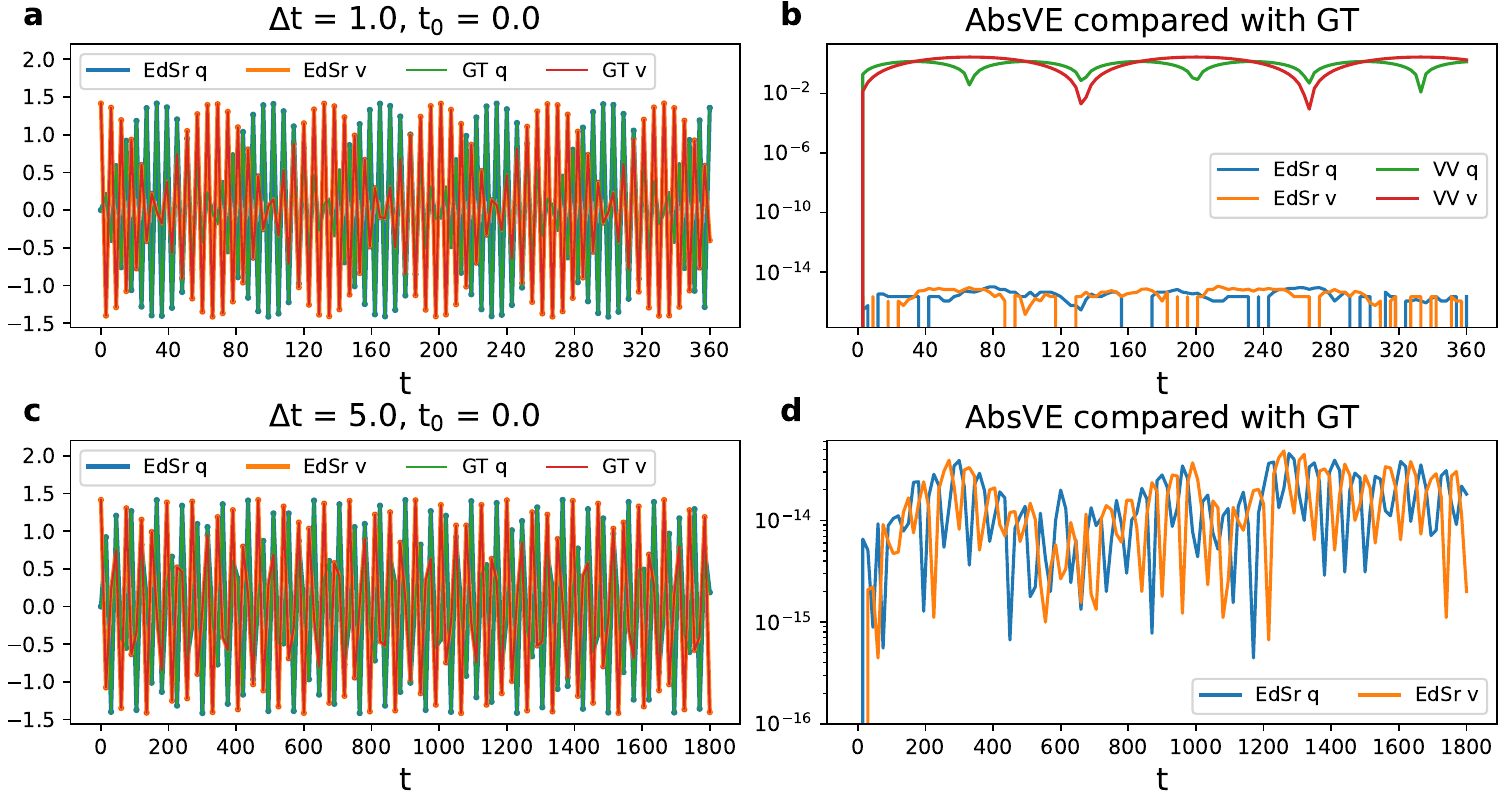}
  \caption{
    Results of the second group of sub-experiments of ideal spring with different \(\Delta t\) values. (a), (b) show the comparisons of generalized coordinates $q$ and velocities $v$ between EdSr and GT, and the corresponding AbsVEs at \(\Delta t = 1.0\).
    (c), (d) denote the results with \(\Delta t = 5.0\). The AbsVEs between VV and GT are not shown at \(\Delta t = 5.0\) as they are very huge. 
  }
  \label{picture:spring_cur}
\end{figure}

\clearpage

\section{Ideal Pendulum}\label{ideal-pendulum}


\begin{figure}[!ht]
  \centering
  \captionsetup{justification=centering}
  \includegraphics[width=0.75\linewidth]{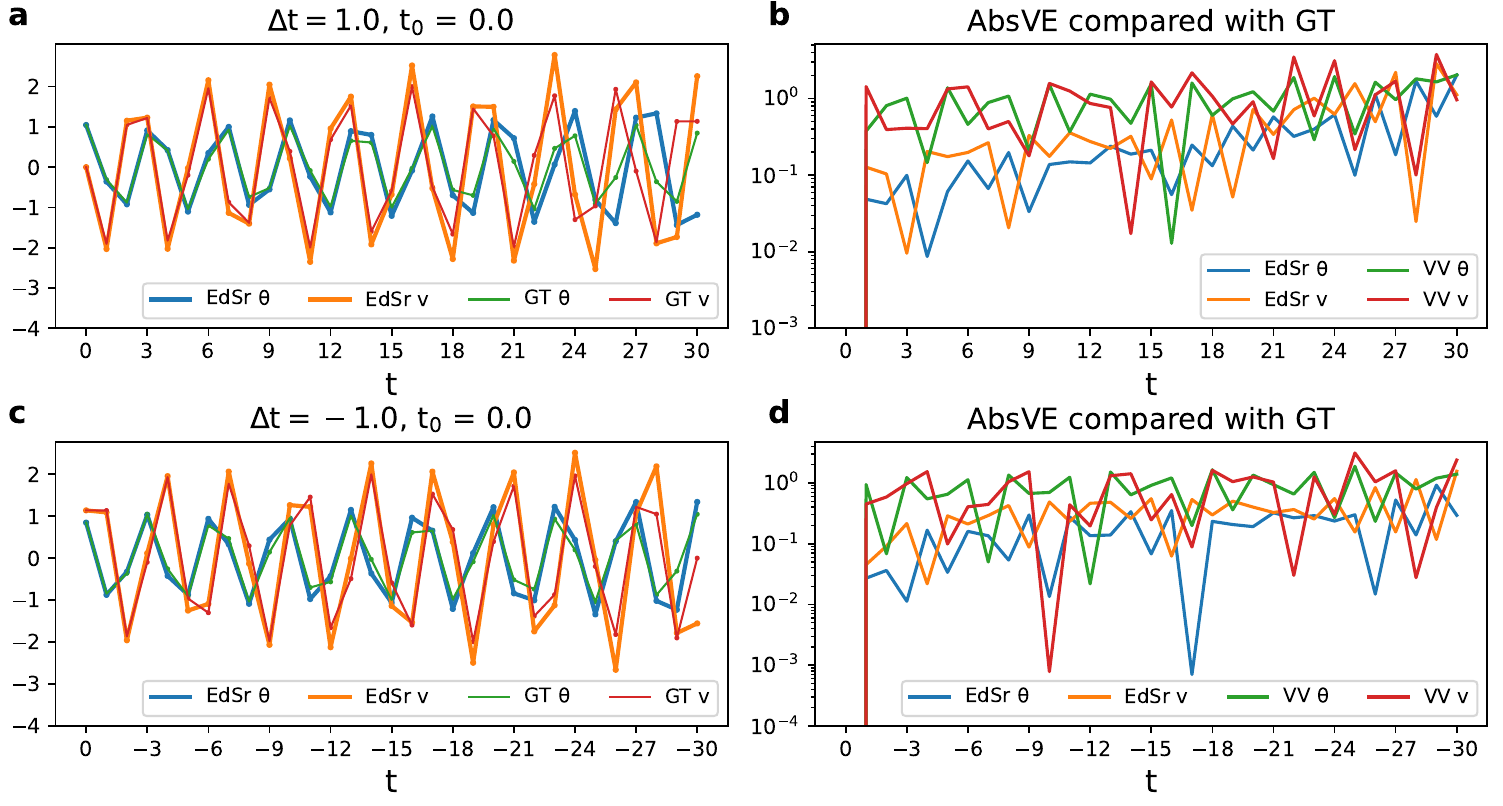}
  \caption{
    Results of the second group of sub-experiments of ideal pendulum with different \(\Delta t\) values. (a), (b) show the comparisons of angles $\theta$ and angular velocities $v$ between EdSr and GT, and the corresponding AbsVEs with \(\Delta t = 1.0\).
    (c), (d) denote the results with \(\Delta t = -1.0\). 
  }
\end{figure}

\newpage

\section{Two Body Model}\label{two-body-model}


\begin{figure}[!ht]
  \centering
  \captionsetup{justification=centering}
  \includegraphics[width=0.9\textwidth]{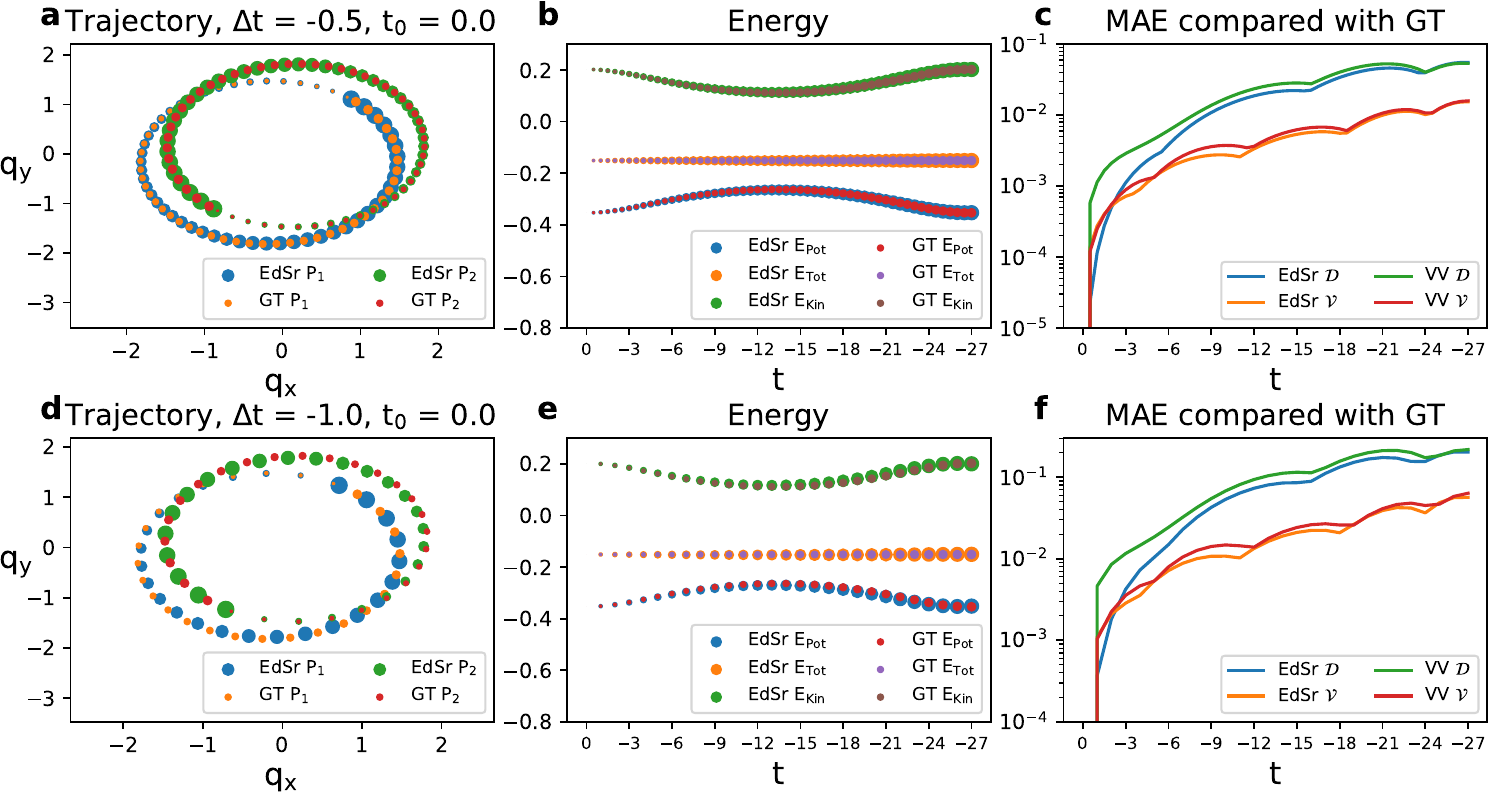}
  \caption{
    Results of the second group of sub-experiments of two-body model at different negative $\Delta t$. 
    (a), (d) denote the time evolutions of positions of the particles in two-body (P$_1$ and P$_2$)  generated by EdSr and GT, the symbol sizes increase as the time starts from 0.0 to -27.0.
    The time evolutions of different energies by EdSr and GT are exhibited in (b), (e).
    (c), (f) denote the MAEs between EdSr and GT, and compare with those between VV and GT over simulation time.
  }
\end{figure}

\begin{figure}[!ht]
  \centering
  \captionsetup{justification=centering}
  \includegraphics[width=0.9\textwidth]{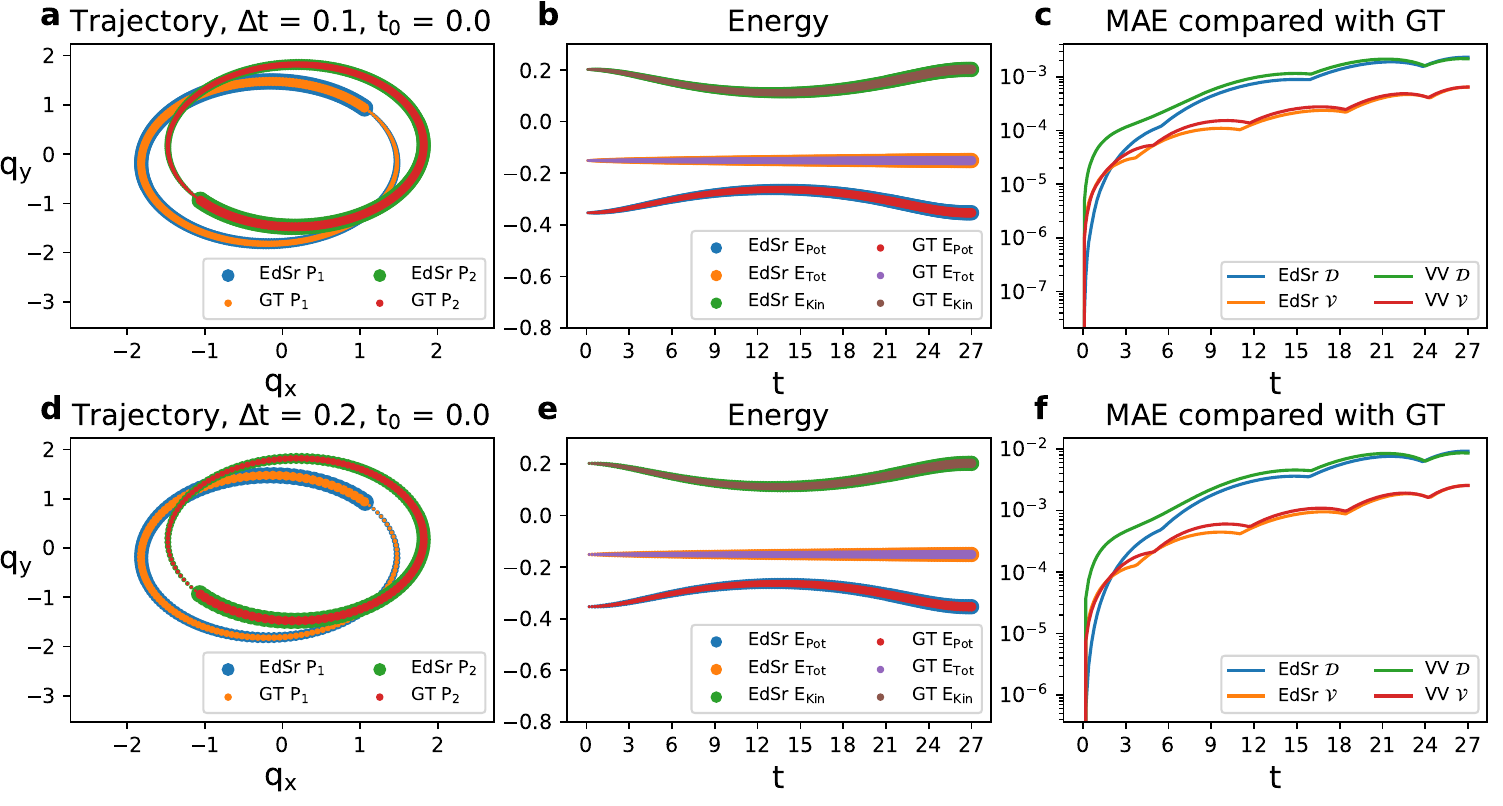}
  \caption{
    Results of the second group of sub-experiment of two-body model at $\Delta t=0.1$ and 0.2. 
    (a), (d) denote the time evolutions of positions of the particles in two-body (P$_1$ and P$_2$)  generated by EdSr and GT, the symbol sizes increase as the time starts from 0.0 to 27.0.
    The time evolutions of energies by EdSr and GT are exhibited in (b), (e).
    (c), (f) denote the MAEs between EdSr and GT, and compare with those between VV and GT over simulation time.
  }
\end{figure}

\clearpage

\section{Diffusion of Indole in Zeolite}\label{indole-in-zeolite}

\begin{figure}[!ht]
  \centering
  \captionsetup{justification=centering}
  \includegraphics[scale=0.75]{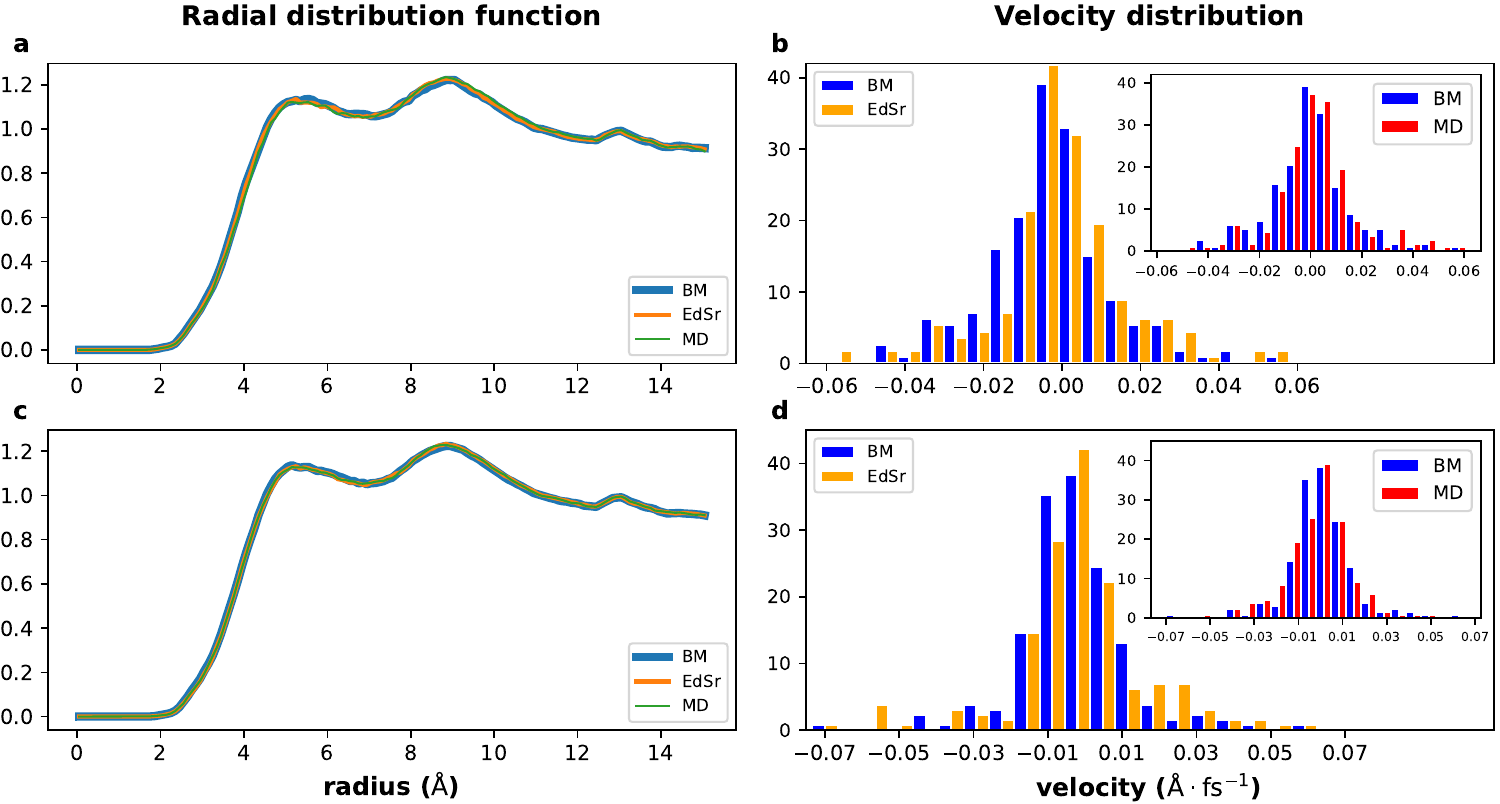}
  \caption{
    Radial distribution functions between indole molecules and zeolite (a) and velocity distributions of all indole molecules (b)  when the simulations are completed at \(\Delta t\) = 1.0 fs.
   (c), (d) show the results at \(\Delta t\) = 3.0 fs.
  }
  \label{picture:indole_rdf_vdist_1.0_3.0}

\end{figure}

\begin{figure}[!t]
  \centering
  \captionsetup{justification=centering}
  \includegraphics[width=1.0\textwidth]{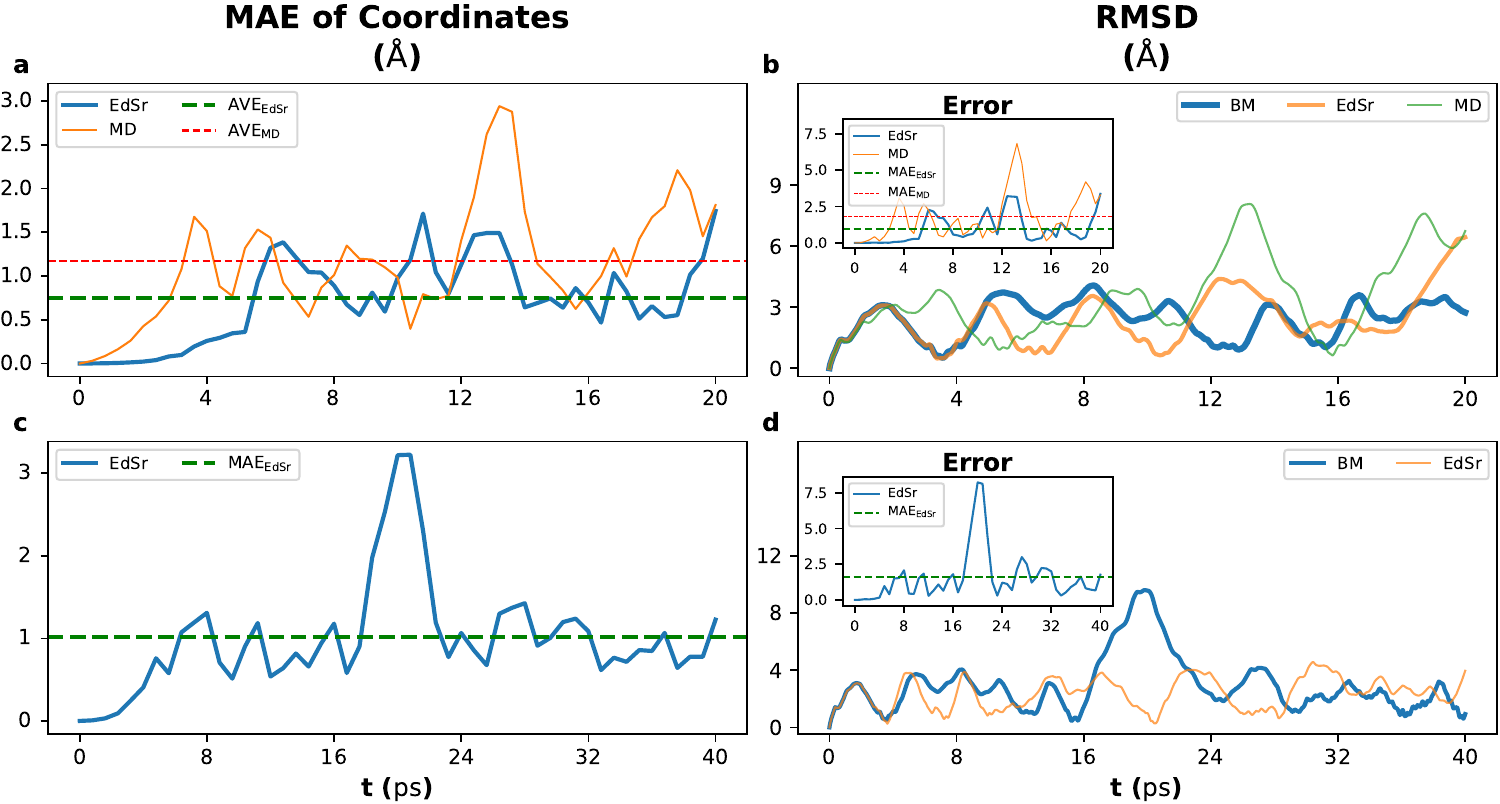}
  \caption{
     MAEs between EdSr and BM groups, as well as between VV and BM groups of the atom coordinates of an indole molecule over time at $\Delta t=2.0$ fs (a) and 4.0 fs (c), the dashed lines in the two sub-figures represent the average MAEs.
    Time evolutions of RMSDs of BM, EdSr and MD groups over time at  $\Delta t=2.0$ fs (b) and 4.0 fs (d), the insets are the AbsVEs, with their MAEs shown as the dashed lines.
    The results at $\Delta t=4.0$ fs are the comparison between EdSr and BM groups, as the MD group can not perform at the timestep.
  }
\end{figure}

\begin{figure}[!ht]
  \centering
  \captionsetup{justification=centering}
  \includegraphics[width=1.0\textwidth]{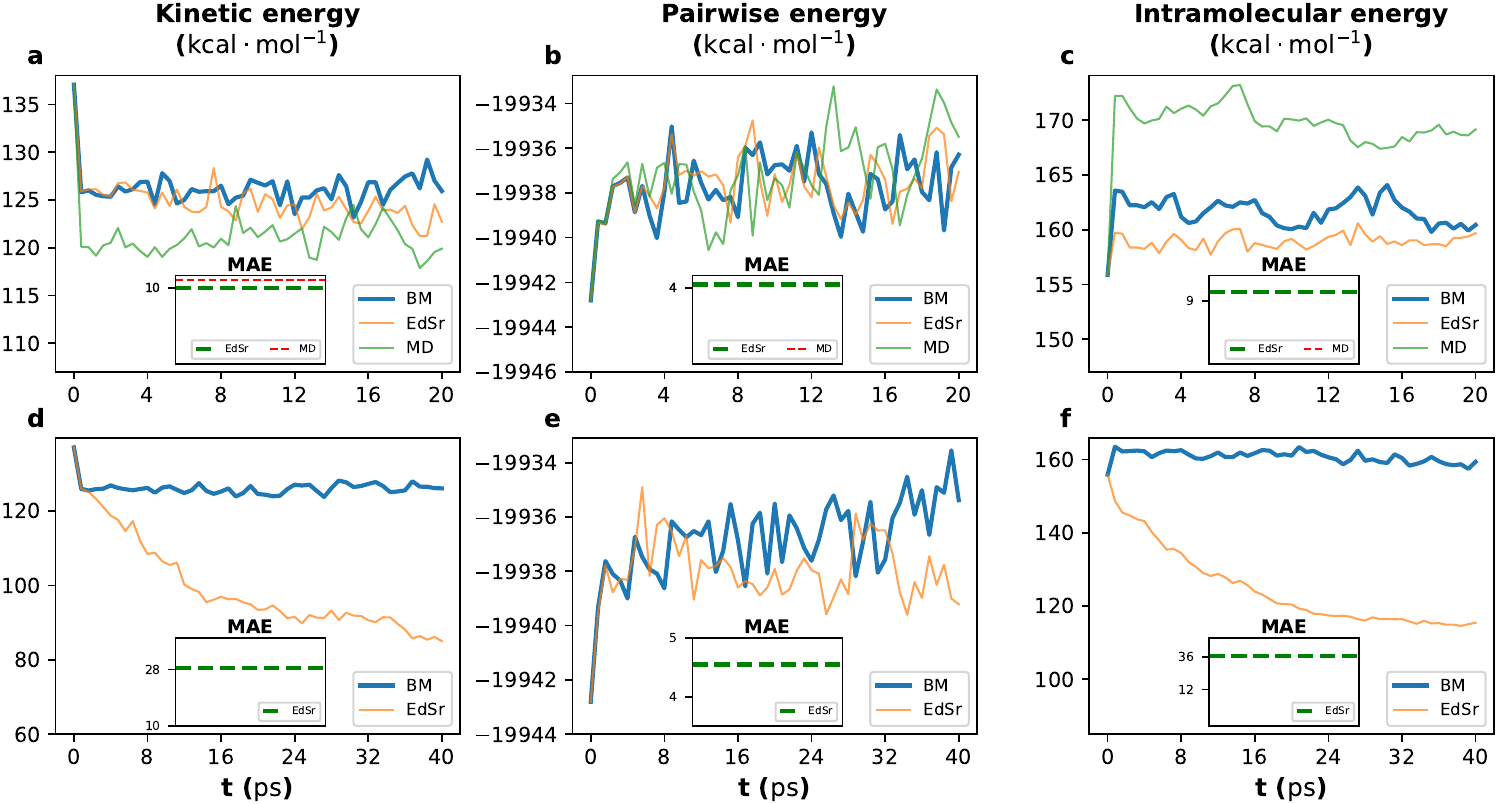}
  \caption{
    Time evolutions of different energies obtained by BM, EdSr and MD groups in all-atom MD simulations. (a-c) show the results of \(\Delta t\) = 2.0 fs,
    (d-f) show the results of \(\Delta t\) = 4.0 fs.
    The MD group can not perform at \(\Delta t\) = 4.0 fs, so there are no  output energies.
  }
\end{figure}

\begin{figure}[!ht]
  \centering
  \captionsetup{justification=centering}
  \includegraphics[width=1.0\textwidth]{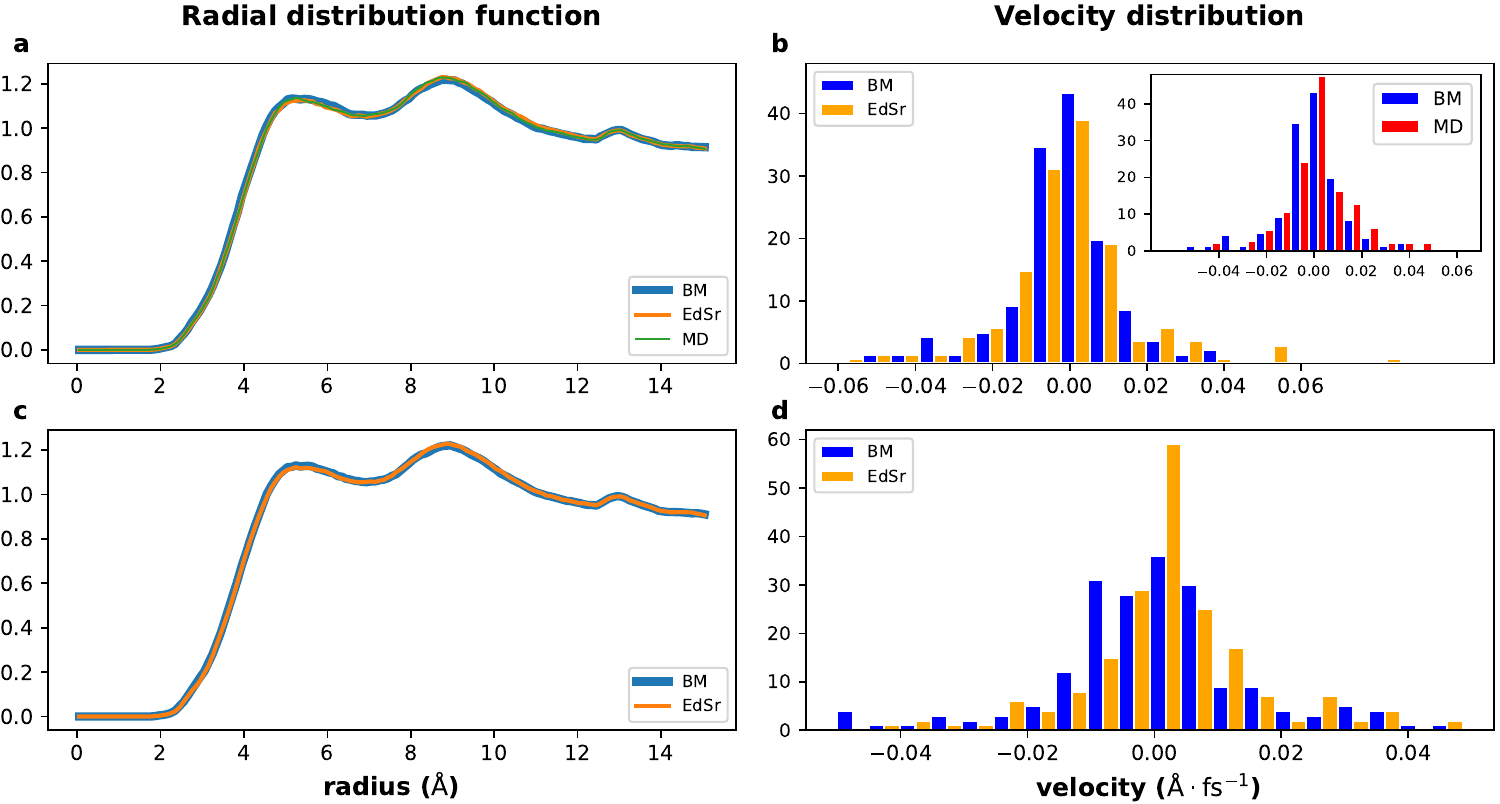}
  \caption{
    Radial distribution functions between indole molecules and zeolite (a) and velocity distributions of all indole molecules (b)  when the simulations are completed at \(\Delta t\) = 2.0 fs.
   (c),(d) show the comparisons between EdSr and BM groups at \(\Delta t\) = 4.0 fs.
  }
\end{figure}

\clearpage

\section{Ubiquitin}\label{Ubiquitin}

\begin{figure}[!ht]
  \centering
  \captionsetup{justification=centering}
  \includegraphics[scale=0.95]{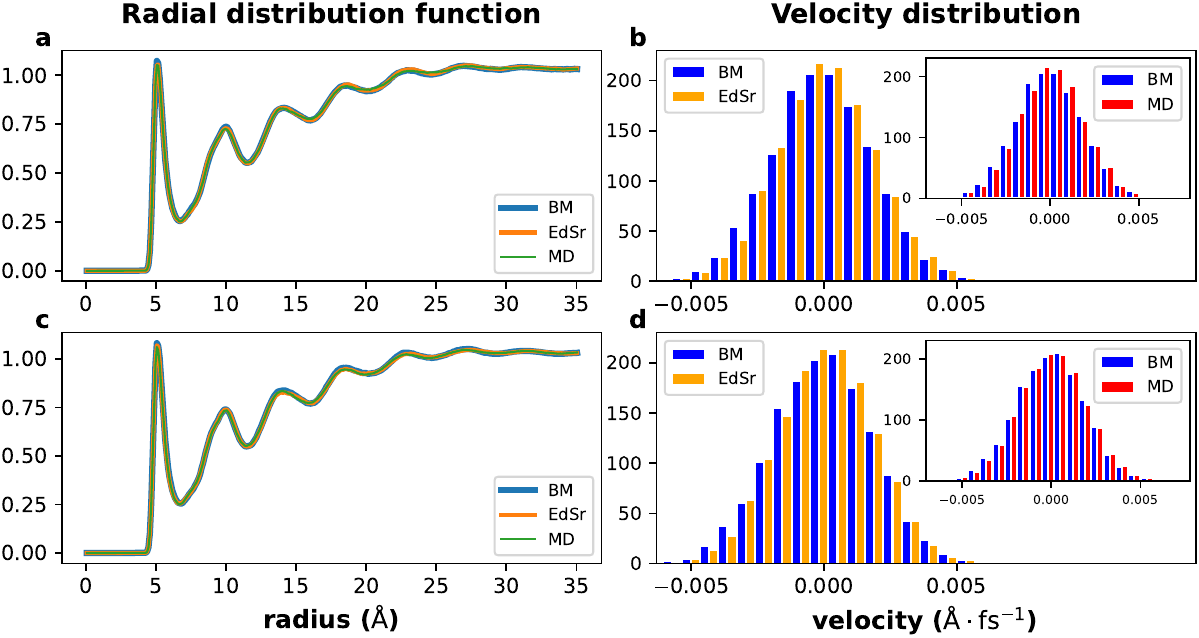}
  \caption{
      Radial distribution functions between ubiquitin and water (a) and velocity distributions of all the particles (b)  when the BM, EdSr and MD groups are completed at \(\Delta t\) = 10.0 fs.
   (c) and (d) show the results at \(\Delta t\) = 20.0 fs.
  } 
  \label{picture:ubiquitin_rdf_vdist_10.0_20.0}
\end{figure}

\begin{figure}[!ht]

  \centering
  \captionsetup{justification=centering}
  \begin{minipage}{1.0\textwidth}
    \includegraphics[trim=0 120 0 0, width=1.0\textwidth]{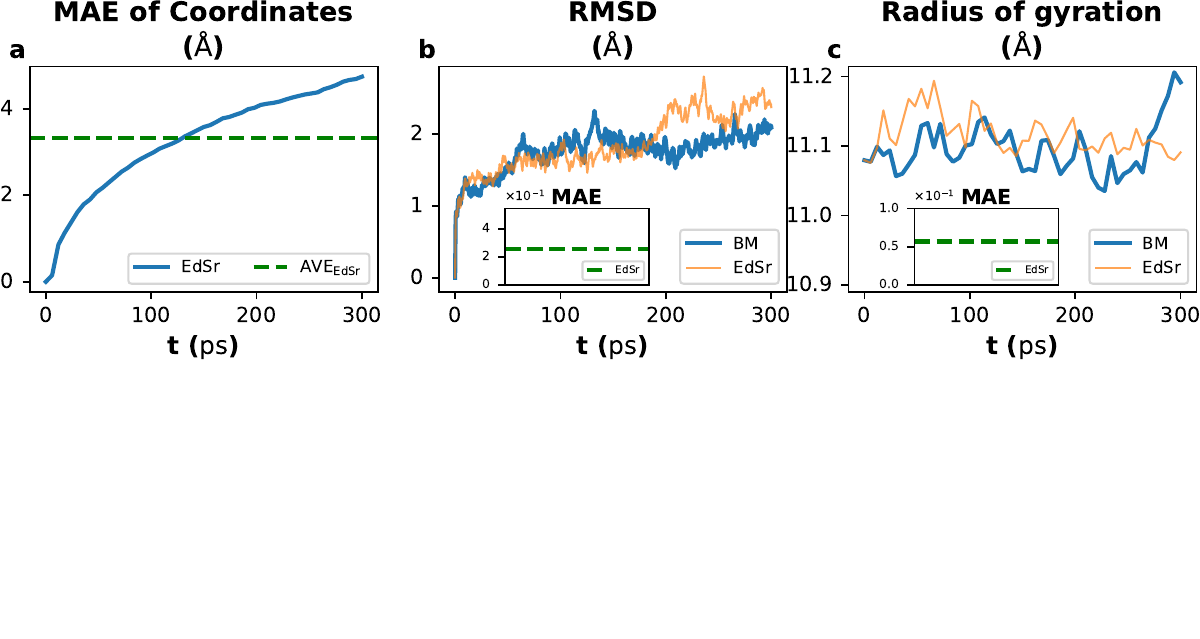}
  \end{minipage}
  \qquad
  \begin{minipage}{1.0\textwidth}
    \includegraphics[trim=0 120 0 0, width=1.0\textwidth]{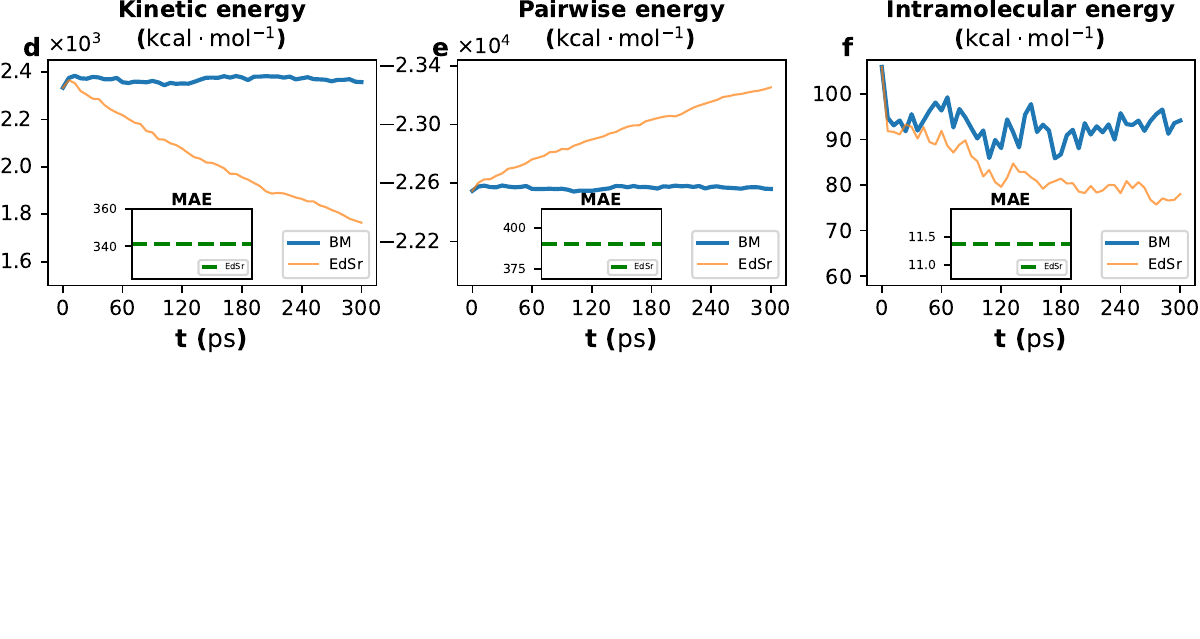}
  \end{minipage}
  \qquad
  \begin{minipage}{1.0\textwidth}
    \includegraphics[trim=0 120 0 0, width=1.0\textwidth]{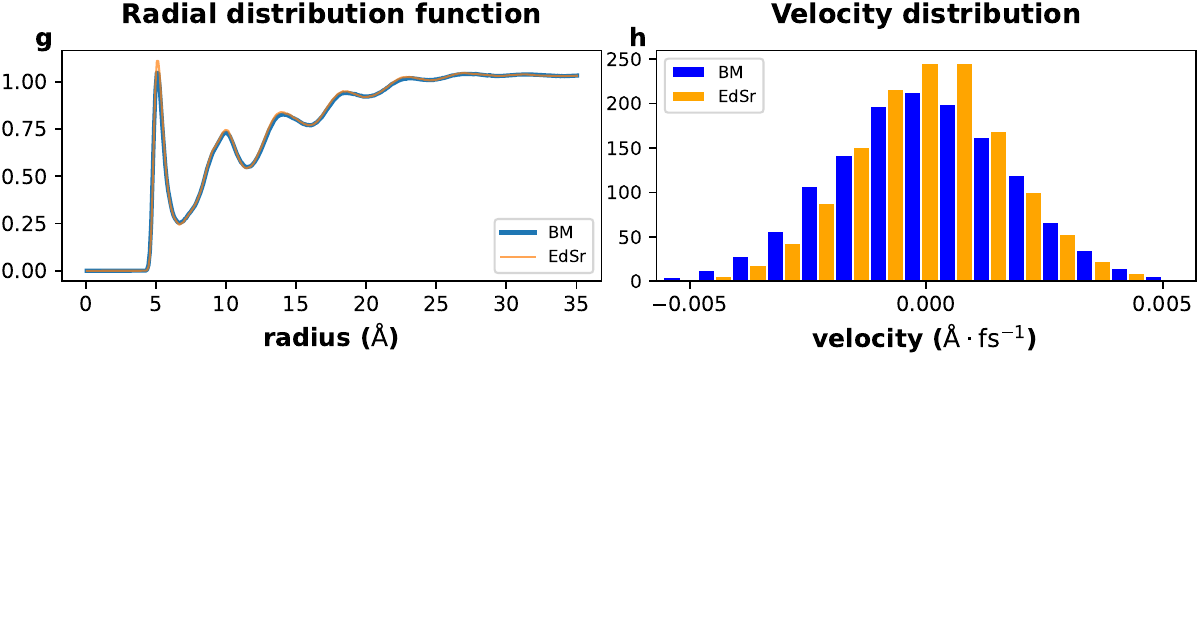}
  \end{minipage}

  \caption{
    Results of ubiquitin at $\Delta t=30.0$ fs by comparing with EdSr and BM groups. 
    The MD group can not perform at such a timestep.
    (a) shows  MAE of coordinates between EdSr and BM. Each point denotes the mean of \(3N\) values, 
    where 3 and $N$ denote dimensions, the number of particles in the system, respectively.
    (b),(c) represent the variation of RMSD, radius of gyration over time, respectively.
    (d-f) denotes the variations of three kinds of energies over time, the insets are the MAEs of EdSr.
    (g),(h) show the results of radial distribution functions, and velocity distributions when the simulations complete.
  }
\end{figure}



\clearpage

\section{Ubiquitin without Water Beads}\label{Ubiquitin-without-water-beads}

\begin{figure}[!ht]
  \centering
  \captionsetup{justification=centering}
  \includegraphics[width=1.0\textwidth]{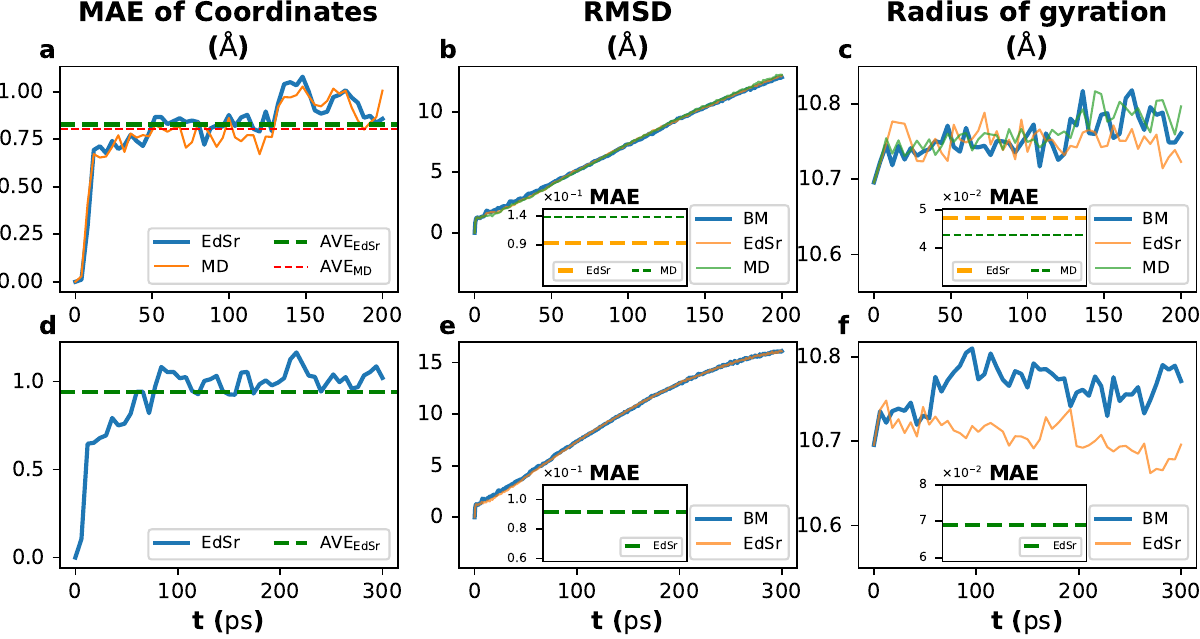}
  \caption{
    The comparisons of MAEs of coordinates, RMSDs, radii of gyration obtained via BM, EdSr and MD groups without water beads.
    (a-c) exhibit the results with \(\Delta t = 20.0\) fs.
    (d-f) exhibit the results with \(\Delta t = 30.0\) fs, the MD group can not perform at \(\Delta t = 30.0\) fs too.
  }
\end{figure}

\begin{figure}[ht]
  \centering
  \captionsetup{justification=centering}
  \includegraphics[width=1.0\textwidth]{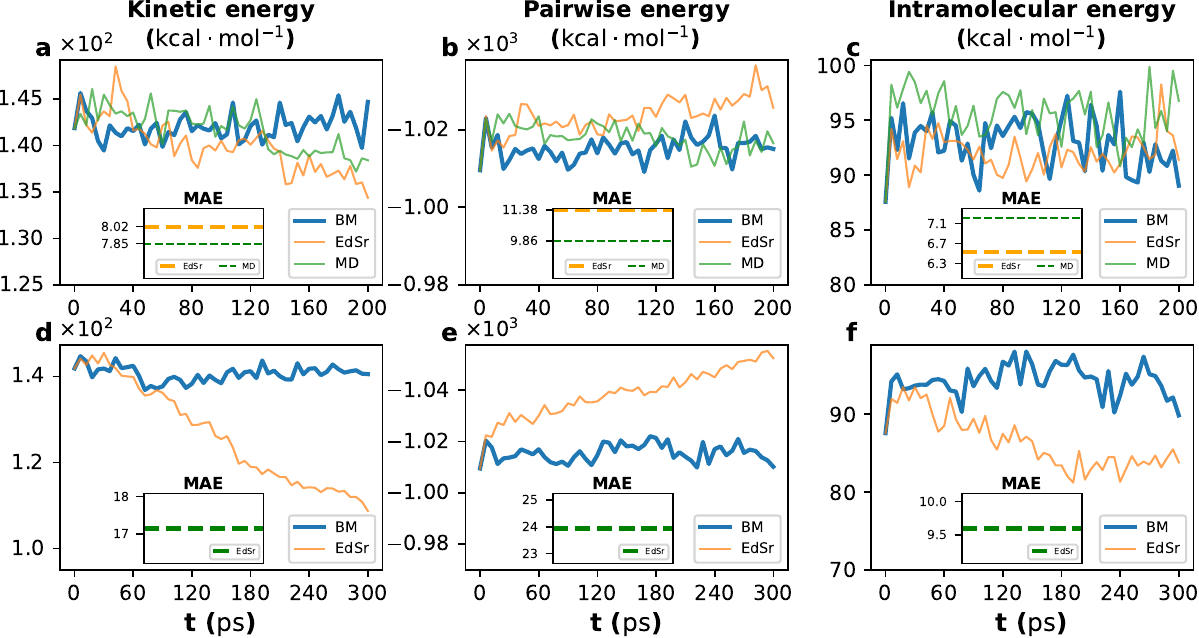}
  \caption{
    Time evolutions of kinetic energies, pairwise energies and intramolecular energies obtained via BM, EdSr and MD groups without water beads.
    (a-c) denote the results with \(\Delta t = 20.0\) fs.
    (d-f) denote the results with \(\Delta t = 30.0\) fs.
  }
\end{figure}

\begin{figure}[ht]
  \centering
  \captionsetup{justification=centering}
  \includegraphics[trim=60 0 70 0, width=0.7\textwidth]{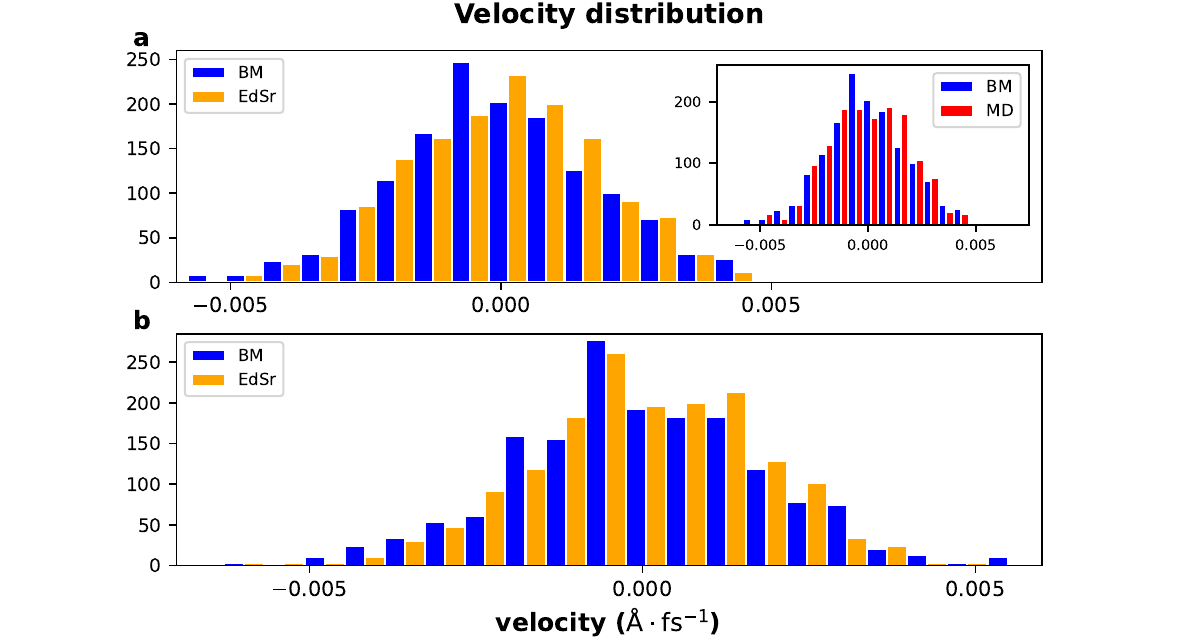}
  \caption{
    Velocity distributions of three dimensions when simulations without water beads complete.
    (a) and (b) show the results at \(\Delta t = 20.0\) fs, \(\Delta t = 30.0\) fs, respectively. 
  }
\end{figure}

\clearpage

\section{Maximum Iteration Time N}
\begin{figure}[h!]
  \centering
  \captionsetup{justification=centering}
  \includegraphics[scale=0.7]{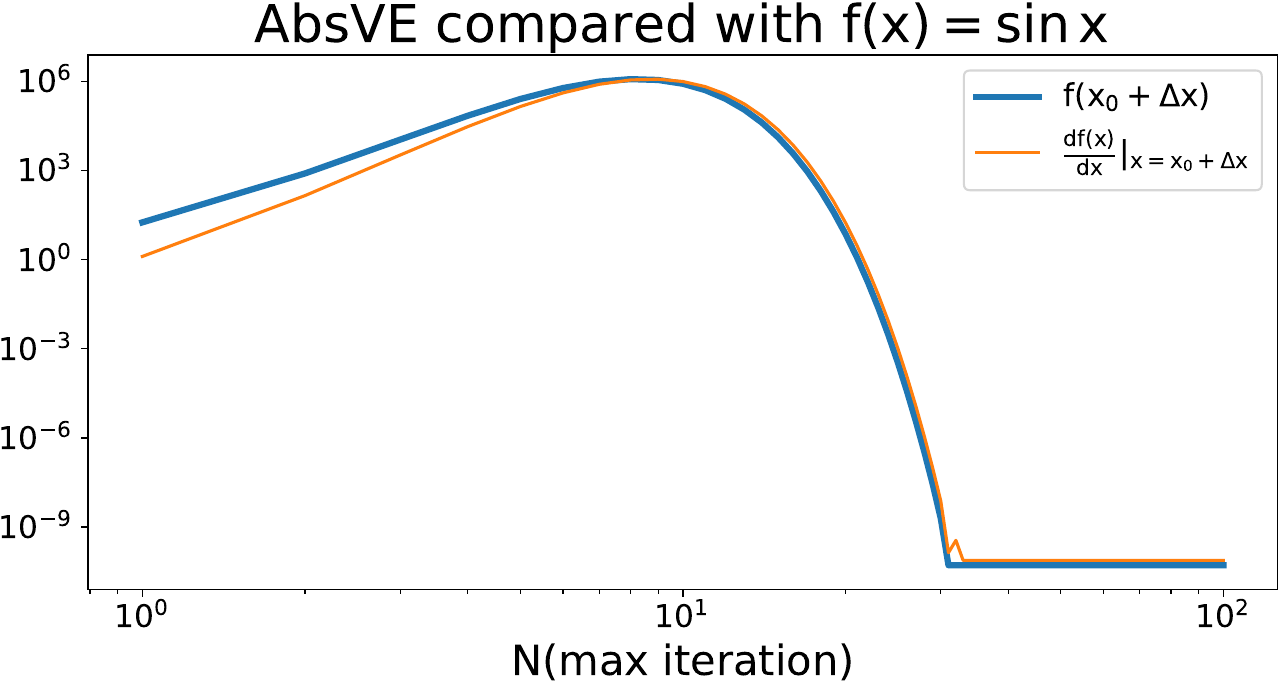}
  \caption{
AbsVEs between the results generated by EdSr with different hyperparameter $N$, and the original function  in the experiment of \(f(x) = \sin x\). 
    Blue line denotes the result of AbsVE compared with \(f(x) = \sin x\).
    Orange line denotes the result of AbsVE compared with the derivative of \(f(x) = \sin x\).
  }
  \label{picture:sinx_N}
\end{figure}

In these experiments, the choice of the maximum iteration number $N$ inevitably arises as a critical issue. Theoretically, the accuracy of EdSr would improve with a large $N$ hyperparameter.
Actually,  the accuracy will not continue to increase once $N$ exceeds a critical value.
The ``critical value'' of $N$ depends on the specific system (or function) to be solved.
Here we take \(f(x) = \sin x\), two body model and all-atom simulation as examples.
For \(f(x) = \sin x\), we select a point (\(\Delta x = 20.0\)) in the middle of the intervals of \(\Delta x\) used in the main article,
and use AbsVE to compare the original function with the result of EdSr at different $N$ hyperparameters.
The result is shown in Figure \ref{picture:sinx_N}.
It is obvious that when $N$ exceeds approximately  30, the accuracy plateaus.

\begin{figure}[h!]
  \centering
  \captionsetup{justification=centering}
  \includegraphics[scale=0.95]{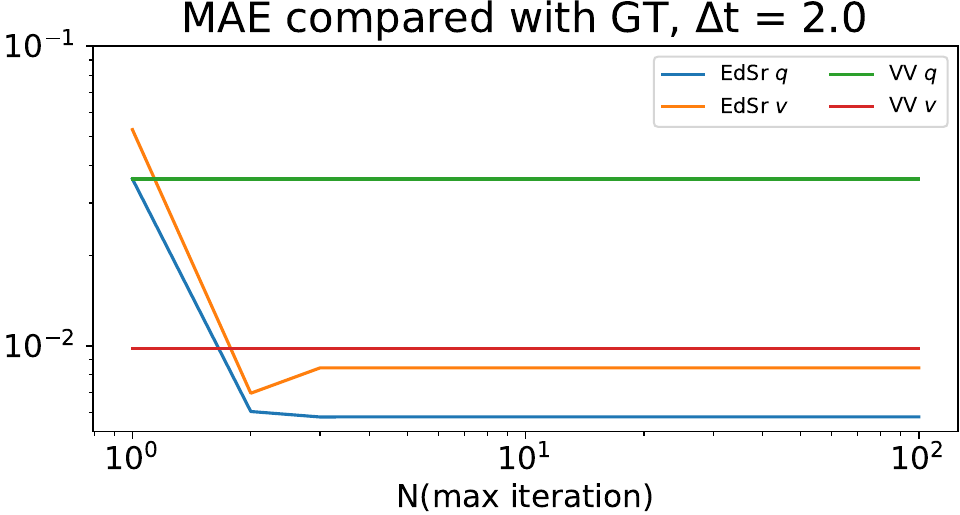}
  \caption{
   MAE results generated by EdSr with different hyperparameter $N$ in the experiment of two-body model. 
    Blue line, orange line denote the results of coordinates, velocities generated by EdSr with different $N$, respectively.
    Green line, red line denote the results of coordinates, velocities generated by VV, respectively.
  }
  \label{picture:twobody_N}
\end{figure}

\begin{figure}[h!]
  \centering
  \captionsetup{justification=centering}
  \includegraphics[width=0.9\textwidth]{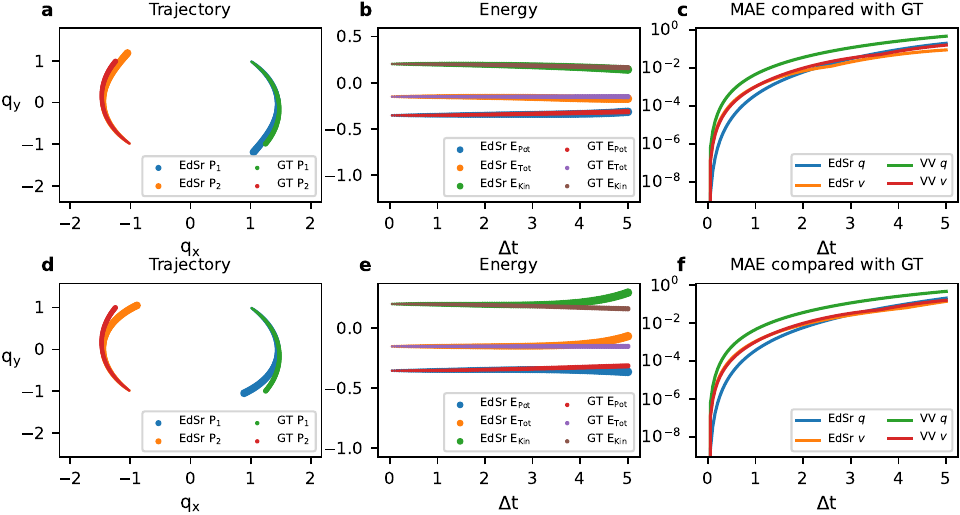}
  \caption{
   The figure shows the comparisons of  EdSr with GT when \(N = 2\)(a-c) and \(N = 4\)(d-f) in the first sub-experiments of two-body Model. 
    (a), (d) denote the positions of the particles in two-body (P$_1$ and P$_2$) generated by EdSr and GT, the symbol sizes increase as the absolute \(\Delta t\) values vary from 0.0 to 5.0. 
    (b), (e) exhibit kinetic energies (\(\mathrm{E_{Kin}}\), green line and brown line), 
    total energies (\(\mathrm{E_{Tot}}\), orange line and purple line),
    potential energies (\(\mathrm{E_{Pot}}\), red line and blue line) from EdSr and GT over \(\Delta t\).
    (c), (f) denote the MAEs between EdSr and GT (blue and orange lines), and compare with that between VV and GT (green and red lines) over \(\Delta t\).
    }
  \label{picture:test_acc_N2_N4}
\end{figure}

Subsequently, for the two body model, 
similar to \(f(x) = \sin x\), we select a fixed \(\Delta t\) (\(\Delta t = 2.0\)) to evaluate the performance of EdSr with different $N$ hyperparameters (shown in Figure \ref{picture:twobody_N}) by countering MAE with GT.
We find that the accuracy converges after only 4 iterations for the two body model, which is significantly fewer than the number of iteration required by \(f(x) = \sin x\).
Furthermore, we show the results of two different hyperparameters of $N$ by setting $N = 2$ (Figure \ref{picture:test_acc_N2_N4}a-c) and $N = 4$ (Figure \ref{picture:test_acc_N2_N4}d-f), which indicates that the computational cost generated by iteration of EdSr does not increase dramatically for some  systems.

\begin{figure}[h]
\centering
\includegraphics[scale=0.9]{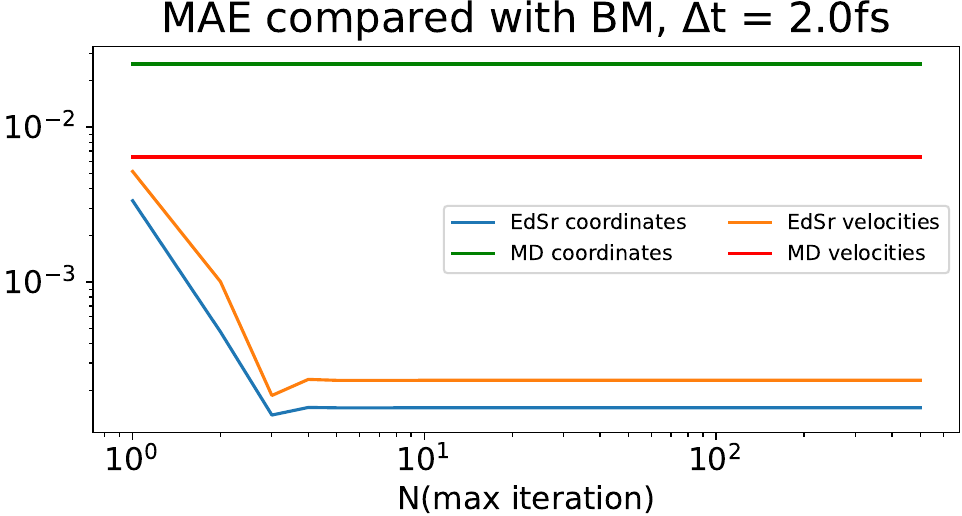}
\caption{    MAEs between the results generated by EdSr and MD groups with comparisons of BM, with different hyperparameter $N$, at $\Delta t=2.0$ fs. 
Blue line, orange line denote the results of coordinates, velocities generated by EdSr with different $N$, respectively.
    Green line, red line denote the results of coordinates, velocities generated in MD group, respectively.}
     \label{picture:all-atom_N}
\end{figure}

We also investigate the performance of EdSr with different $N$ hyperparameters for all-atom simulations at $\Delta t=2.0$ fs.
The results converge after $N=10$ (Figure \ref{picture:all-atom_N}), which is between two-body model and the function $f(x)=\sin x$.

In summary, the selection of the max hyperparameter $N$ involves a trade-off between computational cost and accuracy.
Currently, we  provide two practical strategies for the choice of $N$.
First, we recommend that the turning point to the plateau can be the option for the pursuit of high accuracy (Figure \ref{picture:sinx_N}, \ref{picture:twobody_N} and \ref{picture:all-atom_N}). 
Secondly, if a balance between computational efficiency and accuracy is desired, the $N$ hyperparameters near by tolerance of the specific simulation may  offer a more cost-effective choice. 

\clearpage


\clearpage